\documentclass[CRPHYS,Unicode,manuscript]{cedram}

\usepackage{bm}
\usepackage{soul}
\usepackage{color}
\newcommand{\bfE}{\mathbf{E}}
\newcommand{\bfJ}{\mathbf{J}}
\newcommand{\bfv}{\mathbf{v}}

\graphicspath{{./figures/}}

\title{Laboratory modeling of MHD accretion disks}

\alttitle{Modèles expérimentaux des disques d'accrétion MHD}

\author{\firstname{christophe} \lastname{gissinger}}
\address{LPENS, Ecole Normale Superieure, Paris, France. IUF}
\email[C. Gissinger]{gissinger@ens.fr}
\thanks{The author is supported by funding from the French program ’JCJC’ managed by Agence Nationale de la Recherche (Grant No. ANR 19-CE30-0025-01)  and the Institut Universitaire de France..}


\begin{abstract}
This review article summarizes two decades of laboratory research aimed at understanding the dynamics of accretion disks, with particular emphasis on magnetohydrodynamic experiments involving liquid metals and plasmas. First, the Taylor--Couette experiments demonstrated the generation of magnetorotational instability (MRI) in liquid metals, and highlighted how this instability is critically influenced by boundary conditions and the geometry of the applied magnetic field. These experiments also highlight the nonlinear transition to turbulence in accretion disks, and their link with other MHD instabilities in centrifugally-stable flows. A complementary approach, involving laboratory experiments with volumetric fluid driving rather than rotating boundaries, enables a quantitative study of angular momentum transport by Keplerian turbulence. Collectively, these various laboratory studies offer new constraints on the theoretical models designed to explain the dynamics of accretion disks. This is particularly true with regard to the role of Keplerian turbulence in protoplanetary disks, where recent observations from the ALMA telescope have considerably revised previously expected values of the magnitude of the turbulent fluctuations. Finally, the paper discusses outstanding questions and future prospects in laboratory modeling of accretion disks.
\end{abstract}

\begin{altabstract}
Cet article de revue résume deux décennies de recherche en laboratoire visant à comprendre la dynamique des disques astrophysiques, en mettant particulièrement l'accent sur les expériences magnétohydrodynamiques impliquant des métaux liquides et des plasmas. Tout d'abord, les expériences de Taylor--Couette ont démontré la génération de l'instabilité magnétorotationnelle (MRI) dans les métaux liquides, et ont mis en évidence la façon dont cette instabilité est influencée de manière critique par les conditions aux limites et la géométrie du champ magnétique appliqué. Ces expériences permettent également de mieux comprendre la transition non linéaire vers la turbulence dans les disques d'accrétion et leur lien avec d'autres instabilités dans les écoulements centrifuges stables. Une approche complémentaire, impliquant des expériences de laboratoire avec un entraînement volumique du fluide plutôt qu'avec un dispositif en rotation, permet une étude quantitative du transport du moment cinétique par la turbulence Képlèrienne. Collectivement, ces diverses études de laboratoire offrent de nouvelles contraintes sur les modèles théoriques visant à expliquer la dynamique des disques d'accrétion. Ceci est particulièrement vrai en ce qui concerne le rôle de la turbulence Képlèrienne dans les disques proto-planétaires, dont les récentes observations du télescope ALMA ont considérablement révisé les valeurs précédemment attendues pour l'amplitude des fluctuations turbulentes. Enfin, l'article aborde les questions en suspens et les perspectives futures dans la modélisation en laboratoire des disques d'accrétion.
\end{altabstract}
\begin{document}
\maketitle

\section{Introduction}

Astrophysical disks are present in a myriad of cosmic systems, from protoplanetary disks surrounding young stars (the most common) to accretion disks surrounding compact celestial bodies such as white dwarfs, neutron stars and black holes~\cite{Zhao20, Pringle81, Frank02}. These structures are formed by the gravitational collapse of rotating clouds of gas and dust. In binary star systems, for example, accretion disks form when a star captures material from its companion, creating a rapidly rotating gaseous disk. In the most extreme cases, accretion onto black holes fuels the intense luminosity observed in active galactic nuclei and quasars, and can power gamma-ray bursts~\cite{Fabian12}. Given the diversity of these origins, accretion disks displays a wide range of properties, well beyond the scope of this review. But despite this variety, we will see in this review that astrophysical disks share defining characteristics that allow for their study in a generalized manner within idealized laboratory models. The purpose of these experimental studies, though simplified, is to encapsulate the complex interactions of gravitation, hydrodynamics, and magnetic fields that govern most of the dynamics of accretion disks.

Addressing such complex phenomena undeniably calls for a cross-disciplinary approach, with laboratory experiments playing a pivotal role. Nonlinear dynamics and the role of turbulence in accretion disks are probably among the most important outstanding questions: the process by which turbulence arises and whether fully developed turbulence can account for the substantial amounts of gas and matter typically accreted onto the central object are still unclear.
A key component in understanding this process is magnetohydrodynamics (MHD), which describes the behavior of electrically conducting fluids under the influence of magnetic fields, such as the ionized gas of accretion disks. MHD laboratory experiments, by providing large Reynolds numbers, highly nonlinear regimes, and a strong coupling between the flow and the magnetic field, have offered a valuable opportunity to investigate these two questions under conditions relevant to these disks.

In this review, I will first discuss the main constraints derived from observations and theory, as well as the technical limitations that guide the development of laboratory facilities. I will then describe the extent to which MHD experiments conducted over the last two decades have improved our understanding of accretion disks. In particular, we will see how these studies have focused mainly on the primary instabilities of Keplerian disks and their nonlinear characteristics, the mechanisms responsible for turbulence generation, the transport properties of turbulence, angular momentum dynamics and the role of boundary conditions. In the last part, I will discuss some possible perspectives for the next decade in modeling the magnetohydrodynamics of accretion disks.

\section{Observational constraints, theoretical considerations, and experimental strategies}

Studying accretion disks is challenging due to their vast diversity. Among the simplest are protoplanetary disks, where a dense molecular cloud collapses to form a proto-star surrounded by a protoplanetary disk. In contrast, more exotic configurations include cataclysmic variable stars, a binary pair that involves a white dwarf accreting matter from a companion (usually a classical star), and which exhibits dramatic brightness variations due to limit-cycle instabilities~\cite{Warner03}. Active galactic nuclei (AGNs) present another intriguing case, featuring accretion disks surrounded by moving gas clouds and encircled by a torus of gas and dust, with powerful jets often emerging. Several complex challenges arise in the physics of accretion disks. Radiative transfer within these disks requires a correct understanding of opacity and energy transport to model cooling and\linebreak illumination accurately~\cite{shu1991physics}. Chemical processes and dust dynamics play a critical role in thermal stability and opacity, and are not yet fully understood. Disk-planet interactions, such as planet migration and gap formation, further complicate disk dynamics and planetary system architecture. Modeling accretion disks in the laboratory therefore presents significant challenges, requiring substantial simplifications, compromises, and unavoidable choices in the specific physics processes to be investigated. Many characteristics of astrophysical disks cannot be directly replicated in laboratory setups, which are then guided by observational data and theoretical constraints. The following subsections discuss these observational and theoretical constraints relevant to the specific laboratory modelling of the astrophysical fluid dynamics of accretion disks.

\subsection{Hydrodynamical models}

As this review focuses on the hydrodynamic aspects of accretion disks, it is essential to first address how these disks are modeled using hydrodynamic equations. Accretion disks primarily consist of hydrogen, helium, and trace amounts of heavier elements, with dust contributing only a small fraction of the mass (a few $\%$). Even if the interaction between dust and gas remains significant, initial approximations often concentrate solely on gas dynamics. This single-fluid approach can hinder the investigation of important accretion mechanisms, with the streaming instability being the most well-known example~\cite{Youdin05}. In a typical protoplanetary disk, the mean free path $\lambda$ is of the order of a few centimeters, which is much smaller than the typical macroscopic size $L$ of the disk. Consequently, the Knudsen number $Kn = \lambda/L$ is extremely small, indicating that the continuum mechanics described by the Navier--Stokes equations are generally applicable for modeling disk dynamics. The specific hydrodynamic approximation used can vary depending on the type of disk. For example, in accretion disks around white dwarfs, compressibility effects and supersonic shocks are significant and must be considered~\cite{Wu00}. Conversely, in some protoplanetary disks, such effects may be negligible. Regarding magnetohydrodynamics, the ionization state of the disk varies with distance from the central object. The innermost regions or outer parts of the disk are often highly ionized and dominated by plasma, while cosmic rays and stellar X-rays, which ionize only the surface layers, create a non-ionized central region known as the ``dead zone''~\cite{Gammie96}. This review emphasizes MHD aspects of accretion disks but acknowledges that cold disks or dead zones in classical disks would require consideration of non-MHD dynamics.

\subsection{Disk Geometry and Keplerian rotation}

A first crucial observation concerns the geometry of accretion disks: whether they are spiral galaxies, protoplanetary disks imaged by the Atacama Large Millimeter Array (ALMA) \cite{Andrews18} or disks around compact objects, all observations show that accretion disks take the form of relatively thin disks of gas and dust orbiting rapidly around the central object. This a clear consequence of the force balance. Contrary to a star in which the gravitational pull is in balance with tremendous radial pressure forces, the gravitation in a accretion disk, mainly due to the mass of the central object, is rather supported by the centrifugal force, resulting in the formation of a disk-like structure. Balancing this centrifugal force $U_\varphi^2/r$ with the gravitational pull $GM/r^2$ immediately leads to the velocity profile of an accretion disk, $\Omega \sim r^{-3/2}$, where cylindrical coordinates are used here and $\Omega(r)=rU_\varphi$ is the angular rotation of the fluid. This balance is especially accurate for dust, while the pressure gradient produces a slightly sub-Keplerian velocity profile in the gas, a small correction which is often neglected. However, any rotation profile in which the specific angular momentum $r^2\Omega$ increases outward -- as it does in Keplerian disks -- is known to be stable against Rayleigh's centrifugal instability~(Rayleigh, 1917~\cite{Rayleigh17}). This stability poses a significant challenge in identifying a plausible process that could destabilize a Keplerian disk and eventually induce turbulence.

As we will see, reproducing a thin-disk geometry and a Keplerian rotation in the lab is a challenge and is rarely achieved, but it is essential to capture at least the role of the centrifugal force and the main stability properties of this rotation profile. For this reason, most experiments reviewed here are Taylor--Couette devices, consisting of two concentric cylinders that rotate independently and are bounded vertically by two endcaps (rotating or not). For infinitely tall cylinders, the background flow between these cylinders follows the well known Couette profile:
\begin{equation}\label{couette}
\Omega = A + \frac{B}{r^2}
\end{equation}
where the constants are given by $A=(\Omega_2r_2^2-\Omega_1r_1^2)/(r_2^2-r_1^2)$ and $B=r_1^2r_2^2(\Omega_1-\Omega_2)/(r_2^2-r_1^2)$ and depends on the rotation rates of the inner and outer cylinders, $\Omega_1$ and $\Omega_2$, respectively, as well as the radii of the inner and outer cylinders, $r_1$ and $r_2$. Although achieving a Keplerian rotation rate $\Omega \propto r^{-3/2}$ is not feasible in Taylor--Couette devices, such experiments can still operate under conditions below the Rayleigh stability line. If a centrifugally stable flow ($\Omega_1 r_1^2 < \Omega_2 r_2^2$) is maintained but exhibits an outwardly decreasing angular rotation ($\Omega_1 > \Omega_2$), the flow is termed ``quasi-Keplerian''. it is believed that quasi-Keplerian flows have characteristics quite similar to keplerian disks, particularly with regard to the stability to magnetohydrodynamic instabilities, which are the focus of most MHD experiments.

\subsection{The problem of accretion}

In addition to the azimuthal flow, observations also indicate that Keplerian disks experience significant accretion, meaning there is a substantial radial influx of gas and matter towards the central object. Naturally, such radial accretion is always expected in a gas, because of the presence of angular momentum dissipation due to the molecular viscosity of the gas. The main issue -- and one of the most fascinating aspects of accretion disks -- is that accretion rates are generally extremely large, corresponding to a tremendous transport which is virtually impossible to explain with viscous dissipation alone.
According to median values known from observations, a typical protoplanetary disk of size $L\sim 10$ Astronomical Units (1 A.U. $\sim$ 150 millions kilometers) accretes matter at an astonishing rate of $\dot{M}\sim10^{-8} M_{\odot}$.yr$^{-1}$. But with a mean free path of $\lambda\sim 10$ cm and a thermal velocity $V_T\sim 10^3$m.$s^{-1}$, the typical molecular viscosity of such a disk can be estimated to $\nu\sim 10^2$m$^2$.s$^{-1}$, leading to the accretion timescale $t_\nu\sim 10^{15}$ yr. This timescale, vastly longer than the lifetime of a protostellar accretion disk, implies that almost no accretion would occur in a disk governed solely by molecular viscosity. To explain the observed accretion rates, an additional mechanism, such as instabilities or turbulence, must be invoked. A very useful approach was proposed by Shakura and Sunyaev in 1973~\cite{Shakura73}, where molecular viscosity is conveniently replaced by an effective viscosity induced by small scale turbulent fluctuations:
\begin{equation}
\nu_t=\alpha_{SS}\frac{c_s^2}{\Omega}
\end{equation}
where $c_s$ is the isothermal sound speed, and $\alpha_{SS}<1$ is an adjustable dimensionless parameter, whose precise value depends on the specific mechanism driving the turbulence. As stated above, centrifugal instability is not possible for Rayleigh-stable Keplerian disks, and finding a credible explanation for generating turbulence and large values of $\alpha_{SS}$ in disks has been a central focus of vigorous theoretical investigation over the past few decades.

From an experimental perspective, replicating the strong gravitational field that drives the inward accretion in astrophysical disks is a very complex and challenging task. This might initially seem like a discouraging argument against laboratory experiments. However, fortunately, the conservation of angular momentum dictates that the inward spiraling of matter corresponds to a global outward transfer of angular momentum within the disk: as material moves closer to the central object, it transfers some of its angular momentum to the outer parts of the disk. With appropriately designed rotating experiments (typically Taylor--Couette devices) and precise flow measurements, this flux of angular momentum can be accurately reproduced and studied in laboratory setups, without any need for gravitational infall. We will see that in some cases, some aspects of the gravity field can even be replaced by a similar effect, such as electrostatic~\cite{Hart86} or magnetic forces~\cite{Vernet21a,Vernet22}.

\subsection{Turbulence and angular momentum transport}

In recent years, the advent of state-of-the-art telescopes, such as the Atacama Large Millimeter/submillimeter Array (ALMA) has revealed astonishing substructures in protoplanetary disks and largely reshaped our understanding of turbulence in astrophysical disks~\cite{Andrews18}. For instance, it is now believed~\cite{Flaherty15} that the $\alpha_{SS}$ parameter could be substantially lower than earlier predictions, implying that the expected level of turbulent viscosity might be reduced by several orders of magnitude. This calls into question the relevance of several earlier theories concerning the origin of turbulence and its ability to produce efficient angular momentum transport, such as the MRI discussed in Section~\ref{secMRI}.

Indeed, the theoretical landscape is complex: first, one must identify a sufficiently robust mechanism capable of triggering instability within a centrifugally stable Keplerian flow. Second, the nonlinear development of this instability must lead to turbulence under conditions that are sufficiently universal to accommodate the diversity of observed accretion disks and match constraints of these recent observations. Finally, this turbulent state must align with the observed accretion rates -- a theoretical challenge, as the angular momentum transport in rotating flows may depends critically on the type of turbulence generated.

A variety of theories have been proposed in this perspective. Notable among these are several compelling mechanisms, including gravitational instability~\cite{Lin87}, baroclinic instabilities~\cite{Klahr03}, vertical shear instability~\cite{Urpin98}, to name a few.

Magnetic fields offer another interesting possibility for generating turbulence, but require an electrically conducting plasma. Observations consistently show significant ionization across most accretion disks, albeit localized to specific regions excluding the dead zones previously discussed. This favors theories on powerful but local MHD mechanisms stemming from the interplay between magnetic fields and fluid dynamics in some restricted areas of disks (inner and/or outer parts).

Reproducing these MHD effects in experiments requires careful consideration of several key dimensionless numbers. Besides the conventional Reynolds number $Re = UL/\nu$ (where $U$ and $L$ are respectively the typical azimuthal velocity and size of the disk, and $\nu$ is the molecular viscosity), MHD effects are gauged by the magnetic Reynolds number $R_m = UL/\eta$, where $\eta=1/(\mu_0\sigma)$ is the magnetic diffusivity, $\mu_0$ the vacuum magnetic permeability and $\sigma$ the electrical conductivity. This magnetic Reynolds number simply compares induction effects to the magnetic diffusion. The magnetic Prandtl number $P_m = R_m/Re=\nu/\eta$ compares the two diffusivities. In accretion disks, both Reynolds numbers are typically very high, except in regions with weak ionization where induction is negligible and $R_m$ becomes small. To model MHD effects in laboratory experiments, there are essentially three possible types of fluid : plasmas, liquid metals, or electrolytes. The low electrical conductivity of most electrolytes makes them a less favorable choice, as they only allow for small magnetic Reynolds numbers ($R_m<1$). Given the technical challenges associated with driving a rotating plasma, most MHD experiments have opted for liquid metals, typically liquid gallium or one of its alloys, due to the feasibility of the experiments and usability of liquid gallium at room temperature. Table~\ref{table1} summarizes the physical properties and typical dimensionless numbers for the liquid metal laboratory experiments described in the present article. But in liquid metals, achieving large $R_m$ requires large power. For instance, reaching $R_m > 10$ in a $20$ cm experiment using liquid gallium ($\sigma \sim 3 \times 10^{6}$S.m$^{-1}$) requires large rotation speeds, typically in the range $30-50$ Hz, translating to hundreds of kW in power given the large density of gallium ($\rho\sim 6100$kg.m$^{-3}$). In addition, because $P_m \sim 10^{-5}$ in liquid metals, large $R_m$ can only be achieved in extremely turbulent flows. Plasma experiments might offer alternatives at much higher $P_m$, but they introduce their own technical challenges such as plasma confinement, high temperatures, and measurement difficulties, as detailed in Section~\ref{sec_plasma}.

A certain diversity in experimental setups is therefore crucial for addressing the complex problem of accretion disks, as will be shown throughout this review. In the next sections, \linebreak I will focus on various experimental studies aimed at understanding the magnetohydrodynamics within accretion disks. Given that laboratory experiments can realistically achieve fully turbulent flows and provide accurate long-term statistical data, recent observational breakthroughs have highlighted the critical importance of experimental modeling of accretion disks.

\section{Laboratory studies}

\begin{table}[]
\caption{Physical properties and typical dimensionless numbers for the liquid metal experiments used in modeling astrophysical disks.}
\renewcommand{\arraystretch}{1.30}
\begin{tabular}{|l|l|l|l|l|l|}
\hline & \multicolumn{1}{c|}{Princeton} & \multicolumn{1}{c|}{Promise} & \multicolumn{1}{c|}{Maryland} & \multicolumn{1}{c|} {Kepler} \\ \hline Kinematic viscosity $\nu$ [$m^2.s^{-1}$] & $3,7\times10^{-7}$ & $3,7\times10^{-7}$ & $7,3\times10^{-7}$ & $3,7\times10^{-7}$ \\ \hline Electrical conductivity $\sigma$ [S.m$^{-1}$] & $3,3\times10^6$ & $3,3\times10^6$ & $10^7$ & $3,3\times10^6$ \\ \hline Density $\rho$ [kg.m$^{-3}$] & $6,4\times10^3$ & $6,4\times10^3$ & $0,93\times10^3$ & $6,4\times10^3$ \\ \hline Maximum flow velocities $U$ [m.s$^{-1}$] & 15 & $2\times10^{-2}$ & 20 & 2 \\ \hline Applied magnetic field $B_0$ [mT] & 500 & 10 & 200 & 100 \\ \hline Typical radius $L$ [m] & 0,1 & $6\times10^{-2}$ & 0,1 & 0,1 \\ \hline Reynolds number $Re=UL/\nu$ & $4\times 10^6$ & 3000 & $3\times 10^6$ & $5\times 10^5$ \\ \hline Magnetic Reynolds number $Rm=\mu_0\sigma UL$ & 6 & $5\times10^{-3}$ & 20 & $ 0.8$ \\ \hline Lundquist number $S=\mu_0\sigma B_0L/\sqrt{\mu_0\rho}$ & 2,3 & $2,8\times 10^{-2}$ & 8 & 0.5 \\ \hline
\end{tabular}
\label{table1}
\end{table}

\subsection{Stability of Keplerian flows}

The linear stability of Keplerian rotation with respect to the centrifugal instability is a strong argument for seeking alternative mechanisms for destabilizing a disk. It should be noted, however, that Rayleigh's criterion only provides a criterion for linear stability. At the considerable Reynolds numbers reached in accretion disks, it is quite conceivable that linearly stable accretion disks could be subject to a nonlinear transition to turbulence~(Lesur, 2005~\cite{Lesur05}). This question was investigated by several experimental teams in the early 2010s~\cite{Paoletti12,Eckhardt07, Huisman12} that reported the presence of turbulent fluctuations even in the quasi-Keplerian stable regime.

\begin{figure}[tbp]
\includegraphics[width=0.99\linewidth]{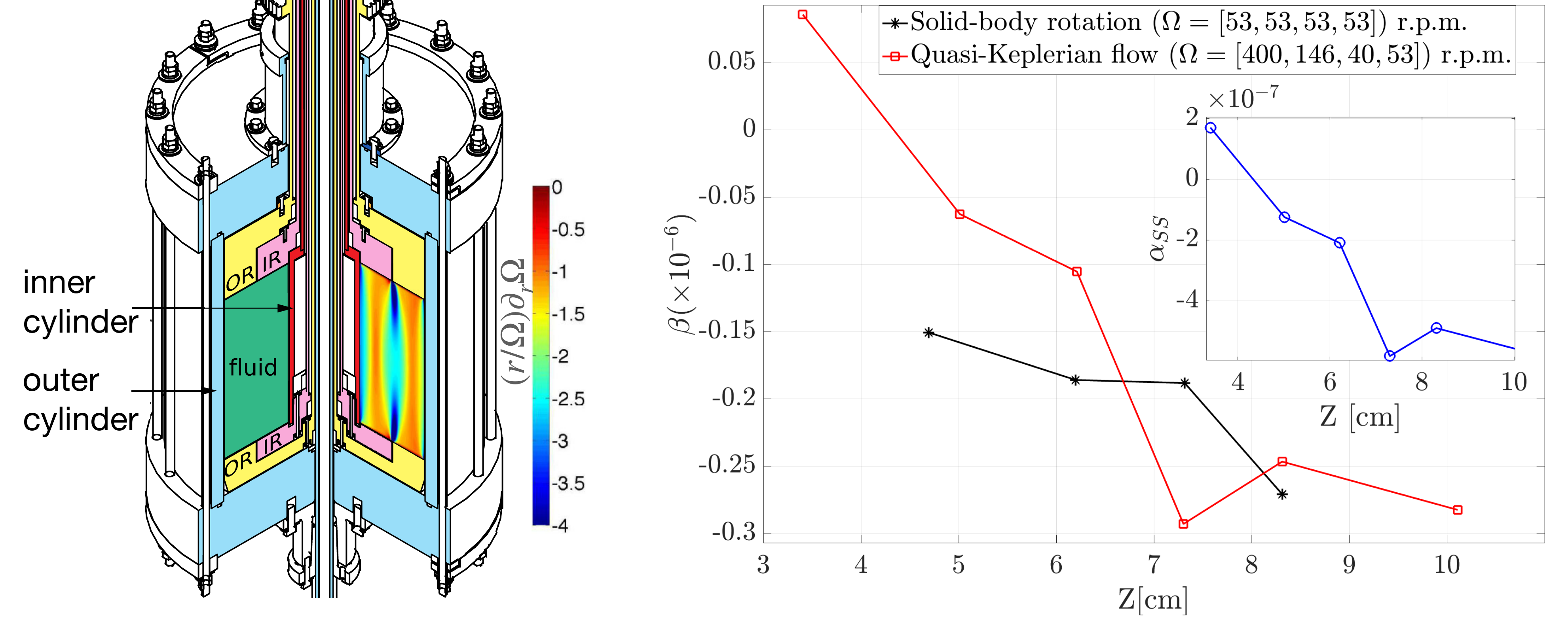}
\caption{Left: Experimental setup used in (Ji et al., 2006~\cite{Ji06}). A rotating fluid (water or a water/glycerol mixture) is confined between two concentric cylinders with a height of $27.86$ cm and radii of $7.06$ cm and $20.3$0 cm, which rotate at rates $\Omega_1$ and $\Omega_2$. Each endcap is divided into two independently driven rings with angular velocity $\Omega_3$ (inner rings) and $\Omega_4$ (outer rings). For appropriate rotations, the secondary circulation is minimized, and ideal Couette profiles are obtained. Right: dimensionless turbulent viscosity $\beta=\overline{u_\varphi' u_r'}/(r^2\partial\overline{\Omega}/\partial r)^2$ measured from velocity fluctuations, for solid-body rotation and quasi-Keplerian flows, showing no sign of transition to turbulence, even at large $Re$ [adapted from~\cite[Figure~3]{Ji06}].
The inset shows the corresponding value of $\alpha_{SS}$ computed for the quasi-Keplerian regime.}\label{fig:Ji06}
\end{figure}

Naturally, it is very tempting to interpret these results as a consequence of boundary effects. Indeed, the viscous friction at the top and bottom endcaps in finite-sized Taylor--Couette devices produces secondary recirculations associated with a strong transport of angular momentum. The velocity profile can deviate from Couette's ideal solution and become unstable, even in the absence of a magnetic field~\cite{Lopez17}. Figure~\ref{fig:Ji06} shows the setup developed at Princeton University to address this issue. The Princeton experiment is a Taylor--Couette device in which the space between two co-axial cylinders is filled with either water or a liquid metal. The two differentially rotating cylinders are confined in the axial direction by two endcaps which, in conventional Taylor--Couette experiments, rotate simultaneously with the inner or the outer cylinder and can drive undesired circulation, as mentioned above. In the Princeton experiment, this issue is solved by dividing each axial endcap into two rings whose rotations can be controlled independently, thus leading to 4 different rotational speeds: the speed of the inner cylinder $\Omega_1$, that of the outer cylinder $\Omega_2$, of the inner rings (upper and lower) $\Omega_3$ and of the outer rings $\Omega_4$. The rotation of the rings can be judiciously chosen to reduce the influence of the system's axial boundaries and get closer to Couette's ideal solution for infinite cylinders. It was shown in~\cite{Ji06,schartman12} that this approach enables hydrodynamically stable flow to be achieved in this geometry at very high Reynolds numbers ($Re > 10^6$). Naturally, this experiment does not definitively eliminate the possibility of a subcritical transition to turbulence in Keplerian flows. However, it demonstrates that Rayleigh-stable Taylor--Couette flows tend to remain stable even at extremely high Reynolds numbers, strongly suggesting the need to identify a more powerful instability to explain the occurrence of turbulence in linearly stable flows.

\subsection{Magnetohydrodynamics and MRI instability}\label{secMRI}

One of the best explanations for the transition to turbulence in such stable flows is to consider the magnetic field. In this case, the traditional framework of hydrodynamics must be replaced by magnetohydrodynamics~\cite{Moffatt78}, accounting for the behavior of the disk's ionized gas, which now acts as an electrically conducting medium. In its simplest form, addressing a dense, incompressible, and homogeneous plasma, the governing equations of MHD are:

\begin{align}
\rho\left(\partial_t \bm{u} +\left(\bm{u}.\bm{\nabla}\right)\bm{u}\right)=-\bm{\nabla} p +\rho\nu\bm{\nabla}^2\bm{u} + \bm{j \times B}\\
\partial_t \bm{B}=\bm{\nabla\times}\left(\bm{u\times B}\right)+\frac{1}{\mu_0\sigma}\bm{\nabla}^2 \bm{B}
\end{align}
where $\bm{u}$, $\bm{B}$ and $\bm{j}=\bm{\nabla\times B}/\mu_0$ are respectively the velocity, the magnetic field and the electrical current of the fluid, and $\mu_0$ the vacuum magnetic permeability. The first equation is the classical Navier--Stokes equation including a Lorentz force which describes the action of the magnetic field on the fluid. For simplicity in this discussion, we assume incompressibility, as most laboratory experiments are conducted at extremely low Mach numbers. However, as mentioned in the introduction and briefly discussed in the conclusion, compressibility effects are expected to play a significant role in some accretion disks. The second equation is called the induction equation and simply corresponds to a combination of Maxwell's equations and the Ohm's law. It describes how the evolution of the magnetic field results from the interplay between induction processes and magnetic diffusion.

A first naive explanation to the accretion of astrophysical disks is that the presence of a magnetic field will produce a torque on the plasma, that can efficiently transport angular momentum. Perhaps the first laboratory experiment aligning with this idea was proposed by~\cite{Donnelly60}, in a Taylor--Couette experiment filled with mercury. It was shown, in agreement with theoretical predictions, that when applied to a rotating flow in a centrifugally unstable regime, a magnetic field sharply increases the onset of centrifugal instability, actually stabilizing the flow. This stabilizing role of the magnetic field, known since the work of Chandrasekhar~\cite{Chandrasekhar_book}, probably explains the lack of interest for MHD processes in accretion disks.

\begin{figure}[tbp]
\includegraphics[width=0.5\linewidth]{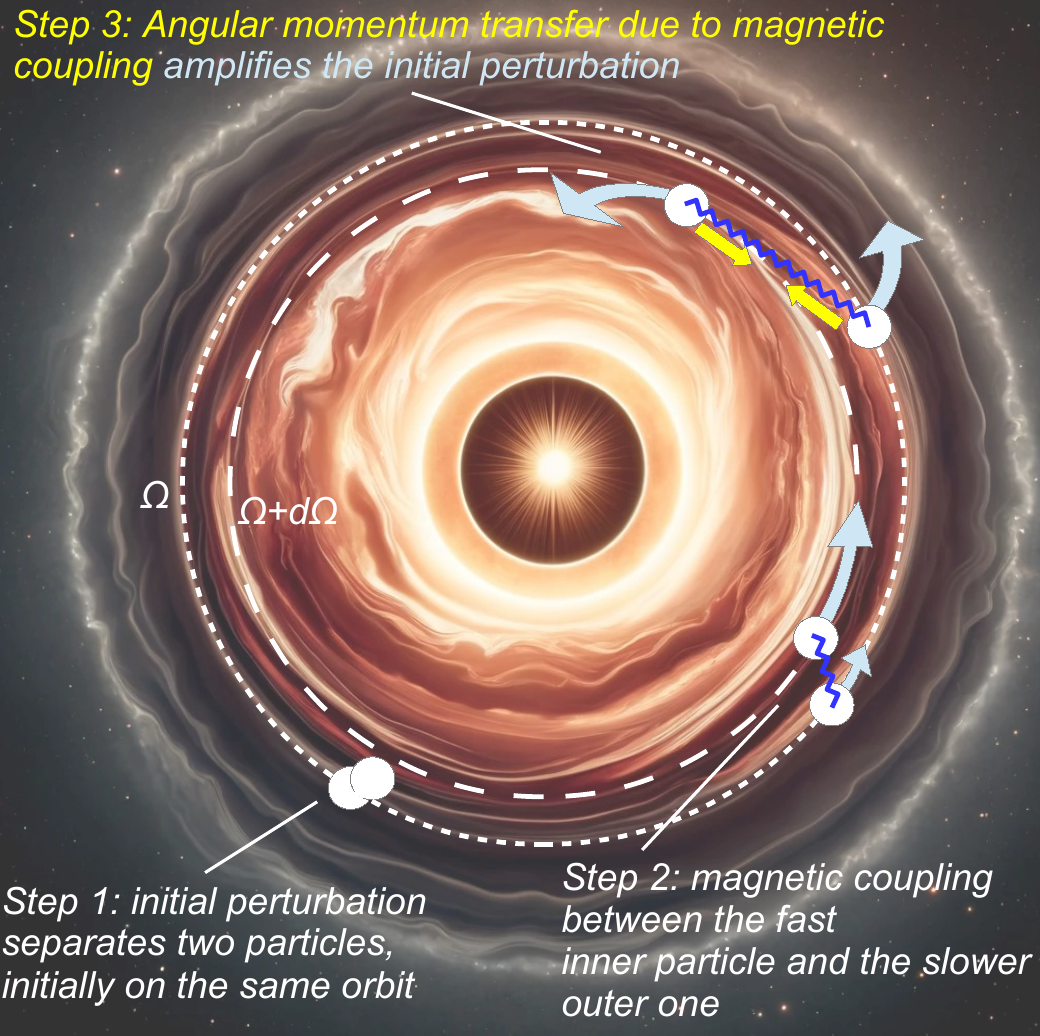}
\caption{Simple sketch of the mechanism for MRI, using the spring model. An initial disturbance separates two fluid particles, initially in the same orbit. The inner particle ends up on a faster orbit as the velocity decreases outwards in a Keplerian flow. The presence of the magnetic field acts like a spring connecting the particles, slowing down the inner particle and accelerating the outer one. This net transfer of angular momentum causes the inner particle to fall a little further inwards, while the other gains angular momentum and moves outwards. The initial disturbance is thus amplified, yielding the flow instability.}\label{fig:mechanism_MRI}
\end{figure}

At least, this was the case until 1991, when a more subtle mechanism was proposed by Balbus and Hawley~\cite{Balbus91}, in which the magnetic field can instead destabilize a centrifugally-stable flow. This is the magnetorotational instability (MRI), which has since become the paradigm for explaining the transition to turbulence in accretion disks. The basic mechanism was initially discovered by Velikhov (1959) \cite{Velikhov59} and Chandrasekhar (1960) \cite{Chandrasekhar60} for Taylor--Couette geometries in a non-astrophysical context, and was later rediscovered as a powerful source of turbulence within accretion disks. The underlying process is elegantly straightforward : the Lorentz force exerted on the fluid divides into two components, $\bm{j\times B}=(\bm{\nabla} B^2+(\bm{B}.\bm{\nabla})\bm{B})/\mu_0$, with the latter often referred to as magnetic tension. This tension acts like an invisible spring, providing a force that binds fluid elements within the disk together. Analogous to how a spring connecting two masses can facilitate the transfer of angular momentum, the magnetic tension generated by the Lorentz force can similarly induce instability in a conducting fluid. This concept is illustrated in Figure~\ref{fig:mechanism_MRI}. This spring-mass analogy for MRI was also reproduced in a clever laboratory experiment that validated many features of the MRI, providing a clear understanding of the physical mechanism behind this MHD instability~\cite{Hung19}.

An important point of the theory is that the instability criterion is remarkably simple and easy to satisfy: any rotation profile where the rotation rate $\Omega$ decreases with increasing distance - as observed in Keplerian disks - will be prone to MRI, even for a weak magnetic field.

It has been shown that this mechanism not only produces a powerful linear destabilization of centrifugally-stable flows, but also that its nonlinear development easily leads to a fully turbulent flow~\cite{Fromang06}. For example, applied to astrophysical conditions, MRI can lead to $\alpha_{SS}\sim [10 ^{-2}-10^{-1}]$, a value large enough to easily explain the highest accretion rates observed~\cite{Fromang06}. Given these considerations, the detection of MRI in large-Re laboratory experiments emerged as an essential undertaking, almost as important as its theoretical demonstration.

In 2001, Goodman and Ji therefore proposed a strategy to detect MRI in a laboratory experiment~\cite{Ji01,Goodman02},which led to the experimental setup described in Figure~\ref{fig:Ji06}. In order to trigger MRI, the device is filled with Galinstan, a room-temperature liquid alloy of gallium, tin, and indium, and subjected to a vertical magnetic field generated by external Helmholtz coils~\cite{Nornberg10}. The experiment can then be operated under quasi-Keplerian rotation, i.e., a flow that is Rayleigh-stable yet potentially MRI-unstable, closely mimicking the dynamics of actual Keplerian disks. Instability within the flow can be detected through various measurements: magnetic perturbations via Hall probes and velocity profiles using ultrasound Doppler velocimetry for instance.

\begin{figure}[tbp]
\includegraphics[width=0.99\linewidth]{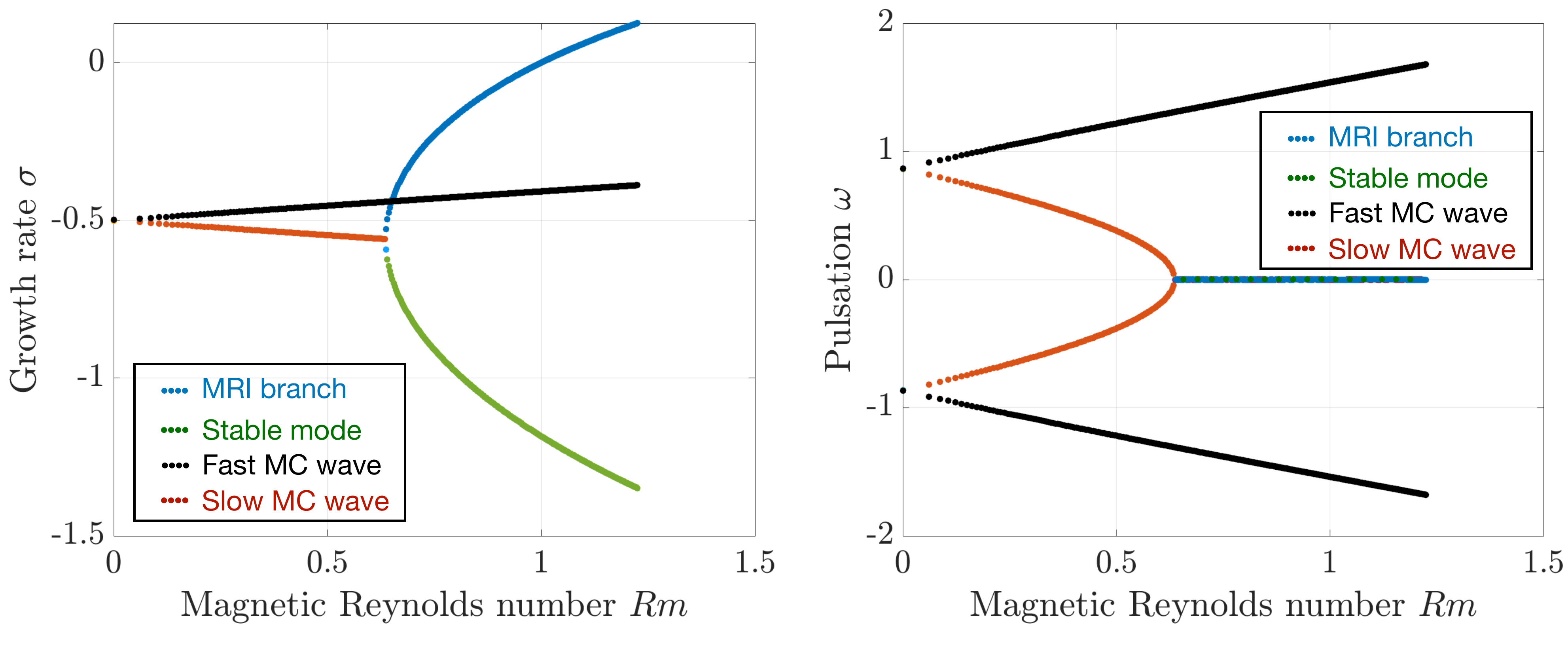}
\caption{ Numerical integration of the dispersion relation~\eqref{dispersion} showing on the left the real part (growth rate) and on the right the imaginary part (pulsation) of the roots of the equation, as a function of $R_m$, for $\varepsilon=1$, $S=1$, $\zeta=0.5$ and $Pm=10 ^{-5}$. At small $R_m$, only damped magneto-coriolis waves are present (red and black curves). But there is a critical value of $R_m$ for which the two slow MC waves coalesce, producing two real eigenvalues from two complex conjugates (blue and green curves). At $Rm=1$ (for the parameter used here), one of these purely real eigenvalues (the blue one) becomes positive, corresponding to the emergence of MRI.}\label{fig:MCwaves}
\end{figure}

This approach proved to be very challenging, as the onset for MRI is difficult to reach in liquid metals. To understand what controls this onset, one can linearize MHD equations around a magnetized Couette solution, and look for infinitesimal perturbations of the form $\exp(\gamma t -ik_zz-ik_rr)$, leading to the classical dispersion relation~\cite{Ji01} for the growth rate $\gamma$:
\begin{equation}\label{dispersion}
\left[(\gamma+P_m)(\gamma+1)+S^2\right]^2\left(1+\varepsilon^2\right) + 2\zeta R_m^2(\gamma+1)^2 -2(2-\zeta)R_m^2S^2=0
\end{equation}
where $\zeta=2+\partial\ln\Omega/\partial \ln r$ characterizes the flow's vorticity and the Rayleigh stability criterion (Rayleigh-stable flows for $\zeta\ge0$), $R_m$ is the magnetic Reynolds number, $\varepsilon=k_r/k_z$, $P_m$ the magnetic Prandtl number and $S=BL/(\mu_0\rho\eta)$ the Lundquist number, a dimensionless measure of the magnetic field. Although highly simplified, this local approach nevertheless captures the essence of the linear aspects of the MRI instability, as shown by the numerical integration of dispersion relation~\eqref{dispersion} in Figure~\ref{fig:MCwaves}. As $R_m$ increases from zero (where rotation effects are negligible and only damped Alfvén waves exist due to magnetic tension), the Alfven solutions split into Magneto--Coriolis (MC) waves (or magneto-inertial waves). These are termed ``slow'' and ``fast'' based on their frequencies and are due to the combined action of magnetic tension and Coriolis force. At a critical $R_m$, complex conjugate slow MC waves merge and form two real eigenvalues (a so-called exceptional point~\cite{Kirillov10}). One of these two branches eventually becomes unstable at high $R_m$: this is the MRI instability, which in this framework, can be simply regarded as a destabilization of magneto-Coriolis waves! This unstable branch also traces back to centrifugal instability if followed into the non-magnetized, Rayleigh-unstable regime.

It is noteworthy that Nornberg et al., 2001~\cite{Nornberg10} reported the first observation of slow magneto-rotational (MC) waves in a Taylor--Couette apparatus designed to detect the magnetorotational instability. The identification of such a precursor to MRI significantly enhances our understanding of how MRI can emerge in laboratory setups.

\subsection{Helical MagnetoRotational Instability and the PROMISE experiment}

In fact, the difficulty of observing MRI in laboratories mainly arises from the high magnetic Reynolds number required, because standard MRI (hereafter called SMRI) necessitates a robust induction process. This led to the clever concept of Helical MRI (HMRI) \cite{Hollerbach05}, where the azimuthal magnetic field, typically induced by diffential rotation at large $R_m$ in SMRI, is now directly applied as part of the experimental setup. In theory, this approach allows for the observation of MRI at much lower rotation rate and magnetic field strengths compared to Standard MRI (SMRI), because this by-passes the requirement of very large rotation rate to induce an azimuthal field.

\begin{figure}[tbp]
\includegraphics[width=0.99\linewidth]{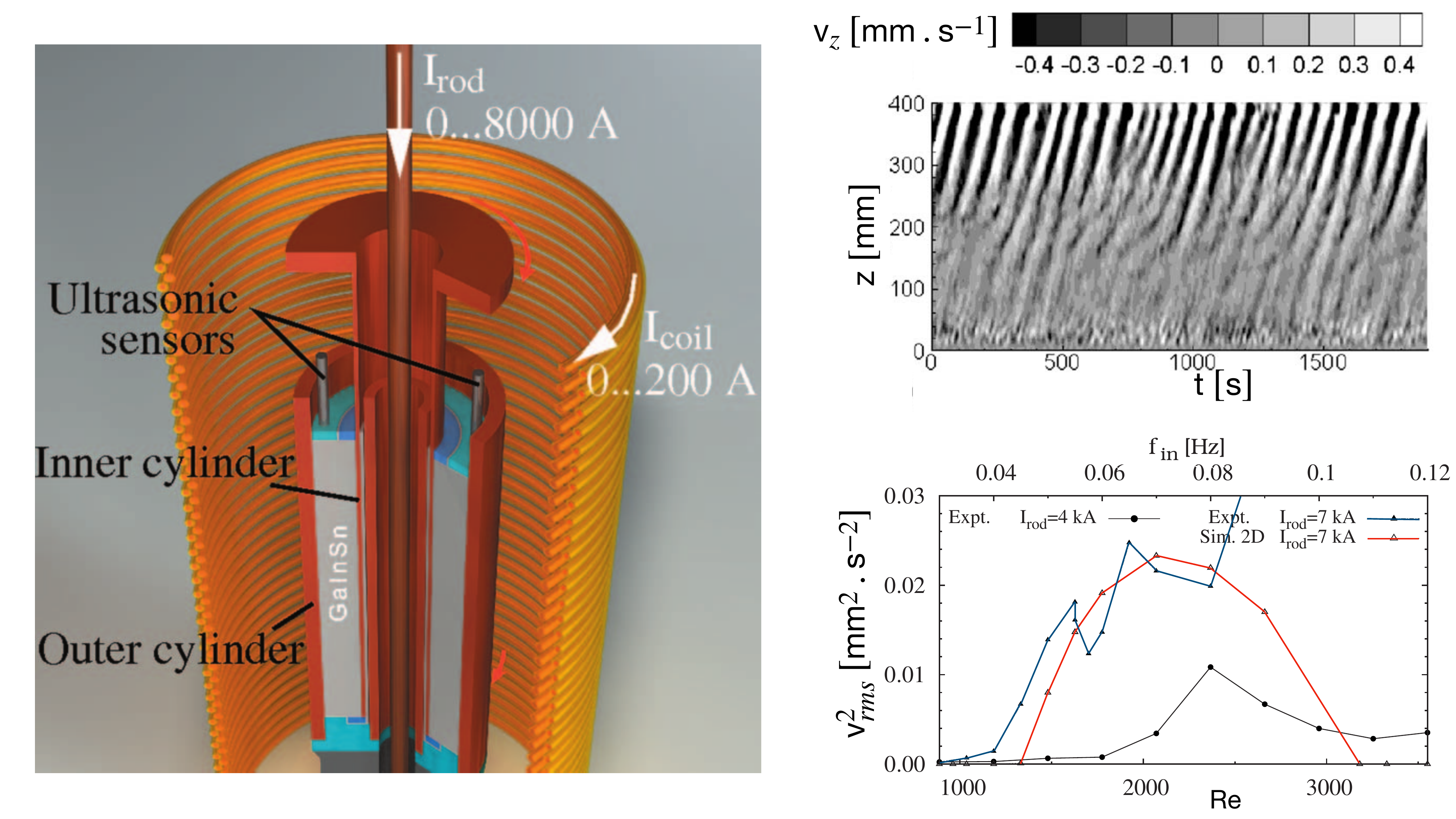}
\caption{Results obtained in the Promise experiment for the Helical MRI~\cite{Stefani09}. Left: experimental setup, with a gap width of $40$ mm and height $400$ mm, showing that the azimuthal field is generated using a very large current passing through a central rod in the inner cylinder. Right, top: spatio-temporal map in the (z,t) plane of the measured axial velocity perturbation $v_z$ obtained from UDV sensors for $Re=3000$, $\Omega_1/\Omega_2=0.27$, $I_{rod}=7000$ A and $I_{coil}=76$ A. The HMRI takes the form of axisymmetric travelling waves. Right, bottom: bifurcation diagram of the MRI obtained from measurements of the rms value of $|\bfv|^2$, displaying an onset close to $Re\sim1000$ and a good agreement with simulations.}
\label{fig:panel_promise}
\end{figure}

The first experimental evidence for HMRI was obtained in 2006 at the PROMISE (Potsdam ROssendorf Magnetic InStability Experiment) facility by F. Stefani and collaborators~\cite{Stefani06}.\linebreak The setup is essentially identical to the Princeton experiment, with the exception of an applied azimuthal field in addition to the traditional vertical field. An interesting destabilization of the otherwise centrifugally-stable flow was reported, taking the form of axisymmetric traveling waves (see Figure~\ref{fig:panel_promise}). HMRI is undoubtedly a new type of magnetorotational instability, but it also fundamentally differs from SMRI: if dispersion relation~\eqref{dispersion} is modified to incorporate the azimuthal field, it shows that HMRI is a magnetic destabilization of pure inertial waves rather than a destabilization of Magneto--Coriolis waves. Consequently, magnetic induction is almost unimportant~\cite{Priede07}, a dynamical regime sometimes referred as inductionless: the induction term $u\times B$ does not need to be significantly larger that magnetic diffusion, and the relevant control parameters is now the Reynolds number $Re$ rather than $R_m$, making the instability onset much easier to reach. Despite this difference, the two instabilities remain connected, as there is a continuous and monotonic transition from HMRI to SMRI when rotation and the magnetic field strength are increased simultaneously~\cite{Kirillov10}. Note that centrifugally-stable Taylor--Couette flows can also undergo a similar instability if the vertical magnetic field is completely suppressed, an instability known as the Azimuthal MRI (AMRI) that was observed a few years later~\cite{Seilmayer14} and may play an important role in stellar interiors.

The application of HMRI to real accretion disks is highly debated. Similar to SMRI, HMRI is expected to produce a fully turbulent flow when operating sufficiently far from the linear onset, particularly at very large Reynolds numbers. Its inductionless nature is extremely appealing, because it provides a simple explanation to turbulence in cold, non-conducting regions of disk, where SMRI cannot operate. On the other hand, it is clear from linear calculations that it only works for relatively steep rotation profiles (i.e., slightly above the Rayleigh line) and disappears in the Keplerian regime~\cite{Liu06}. However, further studies have shown that HMRI can still be relevant in Keplerian flows if at least one radial boundary is highly conducting~\cite{Rudiger07}, a requirement that could be supported by the fact that hot inner part of the disks could be regarded as such conducting boundaries for the colder outer part of accretion disks~\cite{Balbus08}. This is one of many examples of a recurrent pattern in the laboratory modeling of accretion disks: much of the physics is influenced by the boundaries of the system, as we will see in the next section.

\subsection{Boundary effects in Taylor--Couette devices}

It is obvious that the no-slip boundary conditions in experimental setups differ significantly from those in real accretion disks, which, due to their vast extent and lack of clear boundaries, are relatively immune to boundary effects. In this section, we will see how this issue with boundary conditions poses a substantial challenge in modeling laboratory disks, but has also led to some interesting discoveries.

\begin{figure}[!hbp]
\includegraphics[width=0.99\linewidth]{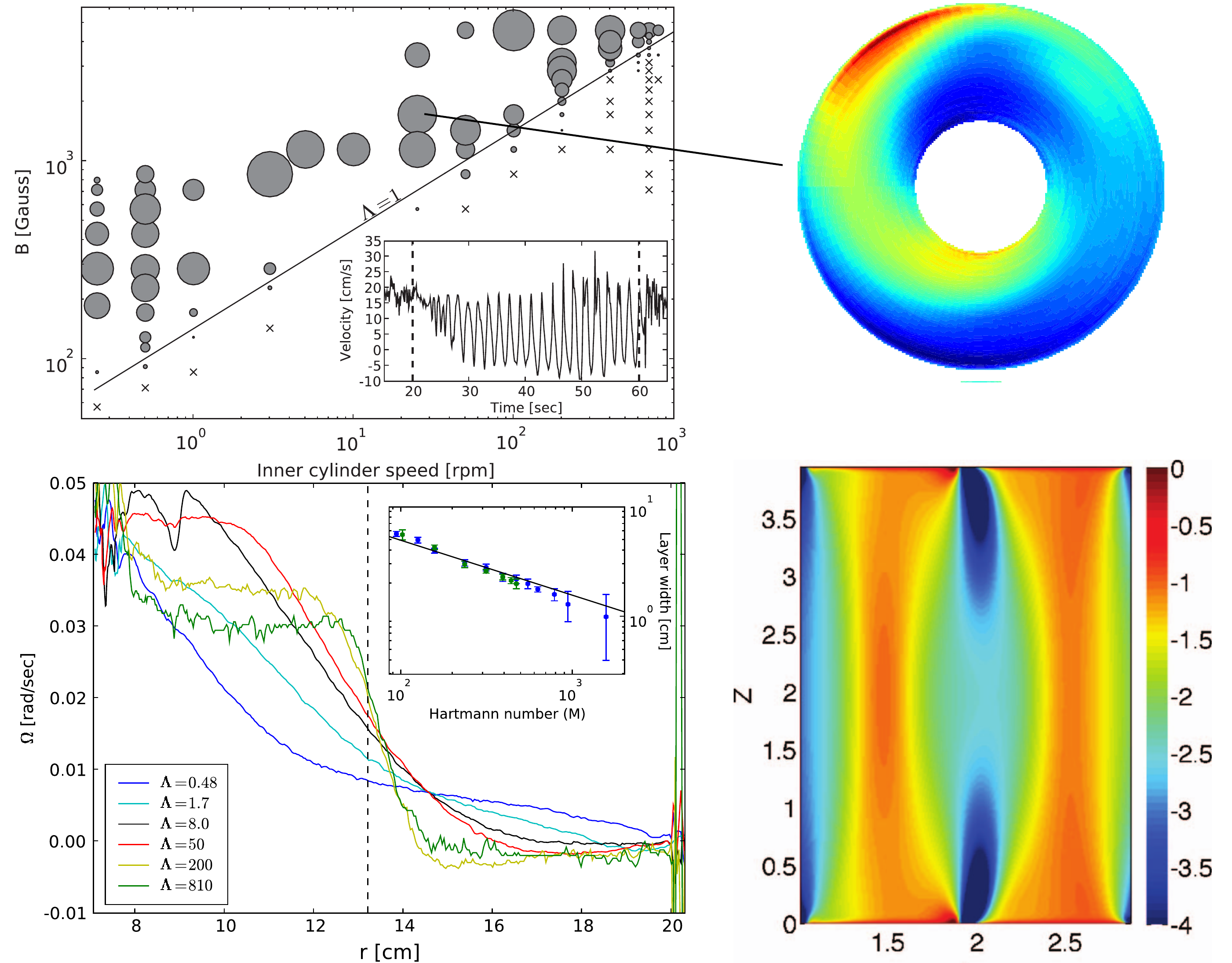}
\caption{Top-left: Stability diagram obtained in~\cite{Roach12} in the ($B,\Omega_1$) plane. The area of the circles is proportional to the power in the dominant Fourier harmonic mode of the instability, while ``x''s indicate stability. The line corresponds to $\Lambda=1$. The inset plot shows a typical time series of the velocity. Top-right: Structure of the unstable mode, reconstructed from Ultrasonic Doppler velocimetry of the azimuthal velocity component (a few cm/s). The spiraling pattern corresponds to an outward transport of angular momentum. Bottom-left: radial profile of the angular rotation $\Omega(r)$ close to the top endcap, for different values of the applied magnetic field (measured by the Elsasser number $\Lambda$). The magnetic field tends to homogenize the flow in the vertical z direction. The velocity jump between the two rings at the endcaps extends into the bulk flow, generating a free shear layer at mid-radius. Bottom-right: snapshot of the unstable eigenmode obtained in the simulations reported in~\cite{Gissinger12}, showing the shear parameter $q(r,z)=r\partial_r\Omega/\Omega$ in the $(r, z)$ plane and illustrating the Kelvin--Helmholtz destabilization of the Shercliff layer.}\label{fig:panel_shercliff}
\end{figure}

The problem of boundary conditions in experimental setups is perfectly illustrated by the role of differentially rotating rings in the Promise and Princeton experiments. This configuration minimizes Ekman recirculation, a necessary condition for efficient detection of the MRI instability. But it was shown in 2012 in the Princeton experiment~\cite{Roach12} that this is also the source of a new class of MHD instability, very close to the MRI but based on a different physical mechanism. Figure~\ref{fig:panel_shercliff} summarizes the results obtained in~\cite{Roach12}: above some critical values of both the magnetic field and the rotation of the cylinders, the flow is destabilized through a supercritical bifurcation of non-axisymmetric modes (azimuthal wavenumbers $m=1$ and $m=2$) which produce a net outward transport of angular momentum. These results are therefore identical to those expected in the case of non-axisymmetric MRI instability. The scaling law is however different. Whereas MRI instability is expected at high $R_m$, flow destabilization in the Princeton experiment occurs systematically for $\Lambda>1$, where $\Lambda=B^2/(4\pi\rho\eta\Delta \Omega)$ is the Elsasser number, and $\Delta\Omega$ is the difference between the inner- and outer-ring rotation rates. This expression interestingly implies that destabilization is possible at very low $R_m$.

Direct numerical simulations (Gissinger et al., 2006~\cite{Gissinger12}) based on this experimental configuration help understanding these confusing results. They show that the difference in velocity between the inner and outer rings generates a flow discontinuity in the middle, which remain localized close to the endcaps as long as the magnetic field is weak. Under a sufficiently strong magnetic field, a Shercliff layer forms, which is distinct from Hartmann layers that are perpendicular to the magnetic field. Classically, Shercliff layers occur near boundaries that are tangential to the magnetic field lines. They result from a local balance between the Lorentz force and viscous friction as the fluid velocity approaches zero. Here however, the Shercliff layer forms due to the velocity discontinuity between the two endcaps' rings, resulting in a free shear layer. Specifically, this layer emerges when magnetic tension stretches the velocity discontinuities into the bulk flow, generating a shear layer characterized by a strong $\partial_ru_\phi$ (see Figure~\ref{fig:panel_shercliff}, bottom-right).
When this free Shercliff layer extends into the bulk, a Kelvin--Helmholtz destabilization of the layer in the azimuthal direction leads to the non-axisymmetric modes reported in the experiment.

\begin{figure}[tbp]
\includegraphics[width=0.6\linewidth]{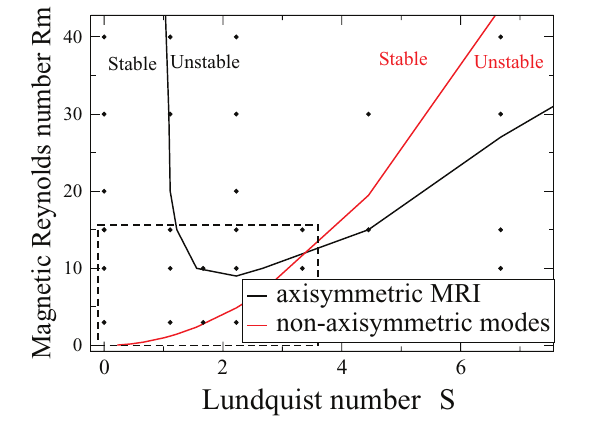}
\caption{Marginal stability curves in the ($R_m,S$) plane computed in numerical simulations of magnetized Taylor--Couette setups~\cite{Gissinger12}. The black curve indicates the stability line of the axisymmetric SMRI. The red curve is the marginal stability for non-axisymmetric Kelvin--Helmholtz instability of the free Shercliff shear layer, obtained from 3D simulations (black dots). The dotted square indicates the parameter space accessible to the Princeton experiment.}
\label{fig:Princeton_shercliff_simu2}
\end{figure}

Figure~\ref{fig:Princeton_shercliff_simu2} illustrates the many similarities between this Shercliff layer instability, and the MRI instability initially sought: in both cases, the instability produces outward transport of angular momentum, requires both a critical magnetic field and a critical rotation rate, and displays a restabilization at high field strength. However, like HMRI, non-axisymmetric Shercliff modes are inductionless, and the relevant dimensionless numbers are different: $Rm$ and $S$ for the MRI, versus $\Lambda$ and $Re$ for this Shercliff layer instability. As with HMRI, the domain of existence of this instability extends to very small $R_m$ (as long as the hydrodynamic Reynolds number $Re$ is sufficiently large).


\subsection{The Maryland experiment}

Of course, it is highly unlikely that such free shear layers play a role in real astrophysical disks. But it is interesting to note that a simple boundary layer instability can produce most of the diagnostics expected for MRI. In fact, it is believed that this scenario is general enough to be observed in other systems, and could explain results obtained in at least one other experimental setup, quite different from the one describe above.

\begin{figure}[tbp]
\includegraphics[width=0.8\linewidth]{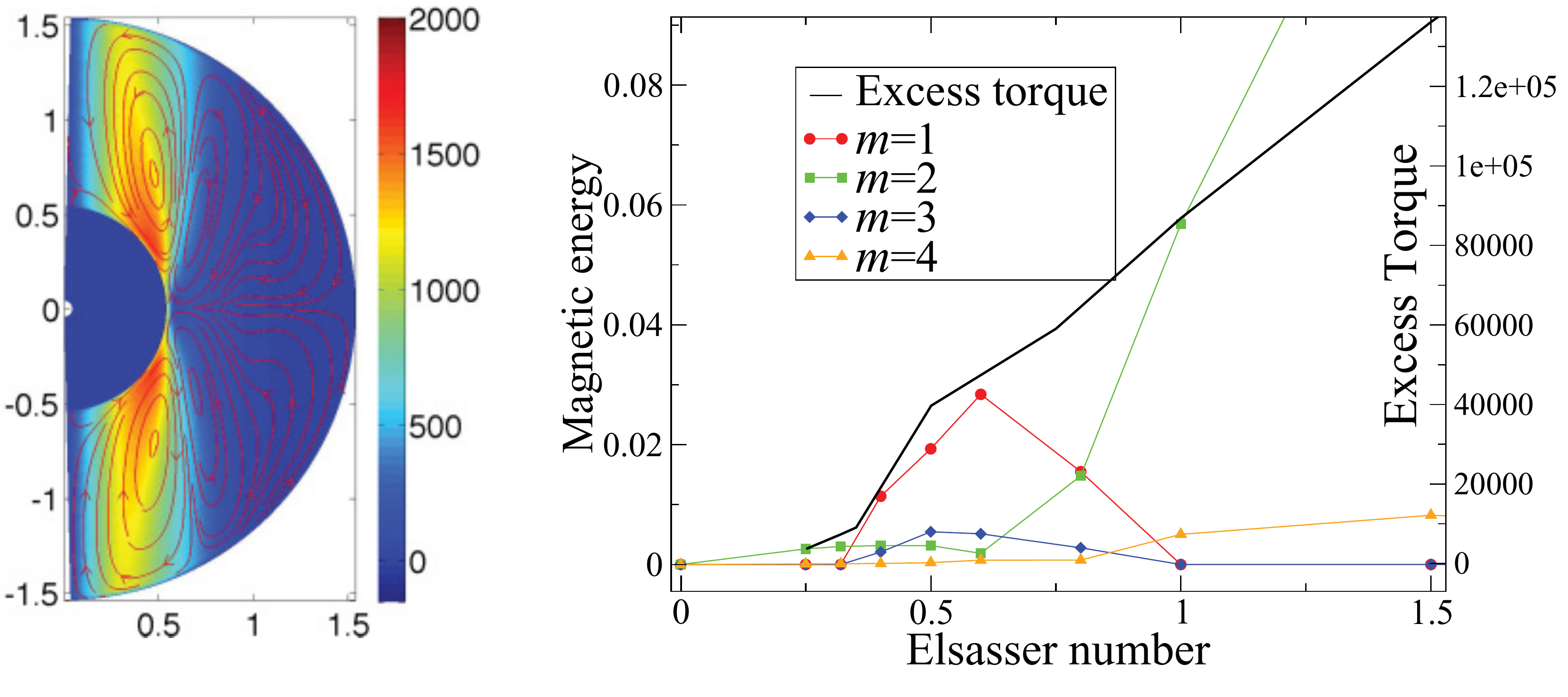}
\caption{Left: structure of the spherical Couette flow when the outer sphere is at rest in a numerical modeling using the same parameters as in the Maryland experiment~\cite{Gissinger11}. Colors indicate azimuthal flow $u_\varphi$ displaying a Shercliff layer surrounding the tangent cylinder. Poloidal field lines of the velocity averaged in the azimuthal direction are also shown. Right: bifurcation diagram of the magnetic energy (contained in different azimuthal wavenumbers $m$) when $\Lambda$ is increased. A good agreement with the Maryland experiment is obtained, including the generation of an $m=1$ mode and increase of the torque on the inner sphere (to be compared with~\cite[Figure~4]{Sisan04}).}\label{fig:simu_Maryland}
\end{figure}

In 2004, the team of Dan Lathrop at the University of Maryland observed an MHD instability in a Couette flow subjected to an intense magnetic field, but this time in spherical geometry~\cite{Sisan04}. In this spherical Couette experiment, where the outer sphere is held at rest and an axial external magnetic field is imposed parallel to the axis of rotation, non-axisymmetric oscillations of the induced magnetic field and velocity were observed, on top of a highly turbulent background flow. These oscillations are associated with a marked increase in torque on the inner sphere, a definitive signature of outward angular momentum transport. A this time, these results were interpreted as an observation of the MRI instability. But numerical simulations of the Maryland experiment~\cite{Gissinger11} have clearly called for a different interpretation in terms of an instability of free shear layers, a priori unrelated to MRI. Indeed, a similar Shercliff layer instability has been numerically predicted in magnetized spherical Couette flows similar to the Maryland experiment, showing striking similarities with the experiment (see Figure~\ref{fig:simu_Maryland}). In this case, instability results from differential rotation between the inner and outer spheres, with the shear layer localized along the tangent cylinder. As in the cylindrical case, the Shercliff layer instability exhibits all the features of MRI, including a considerable increase of angular momentum transport, as shown by the excess torque on the outer sphere. The excellent agreement between these numerical results and those reported by the Maryland group in 2006~\cite{Sisan04} led to a reinterpretation of their experiment as a Shercliff layer instability rather than a detection of the magnetorotational instability.

\subsection{Nonlinear aspects of the MRI}

A general statement must be made here, which applies beyond the physics of accretion disks. In most fluid dynamics experiments, experimental constraints often force researchers to use geometries and boundary conditions that do not match the idealized boundaries considered in theoretical models or the actual boundary conditions found in astrophysical and geophysical systems. It is generally well accepted that boundaries will influence the linear properties of the studied instability, such as wavenumber and threshold. However, in many cases, these boundary modifications can also significantly impact the nonlinear dynamics of the system. This section illustrates this aspect with two examples: first, how boundary-driven recirculation can alter the nature of the flow bifurcation; and second, how it affects the mechanism for the saturation of the underlying instability. Both examples highlight the important and sometimes underestimated connection between the nonlinear dynamics of fluid systems and boundary conditions.

\begin{figure}[tbp]
\includegraphics[width=0.8\linewidth]{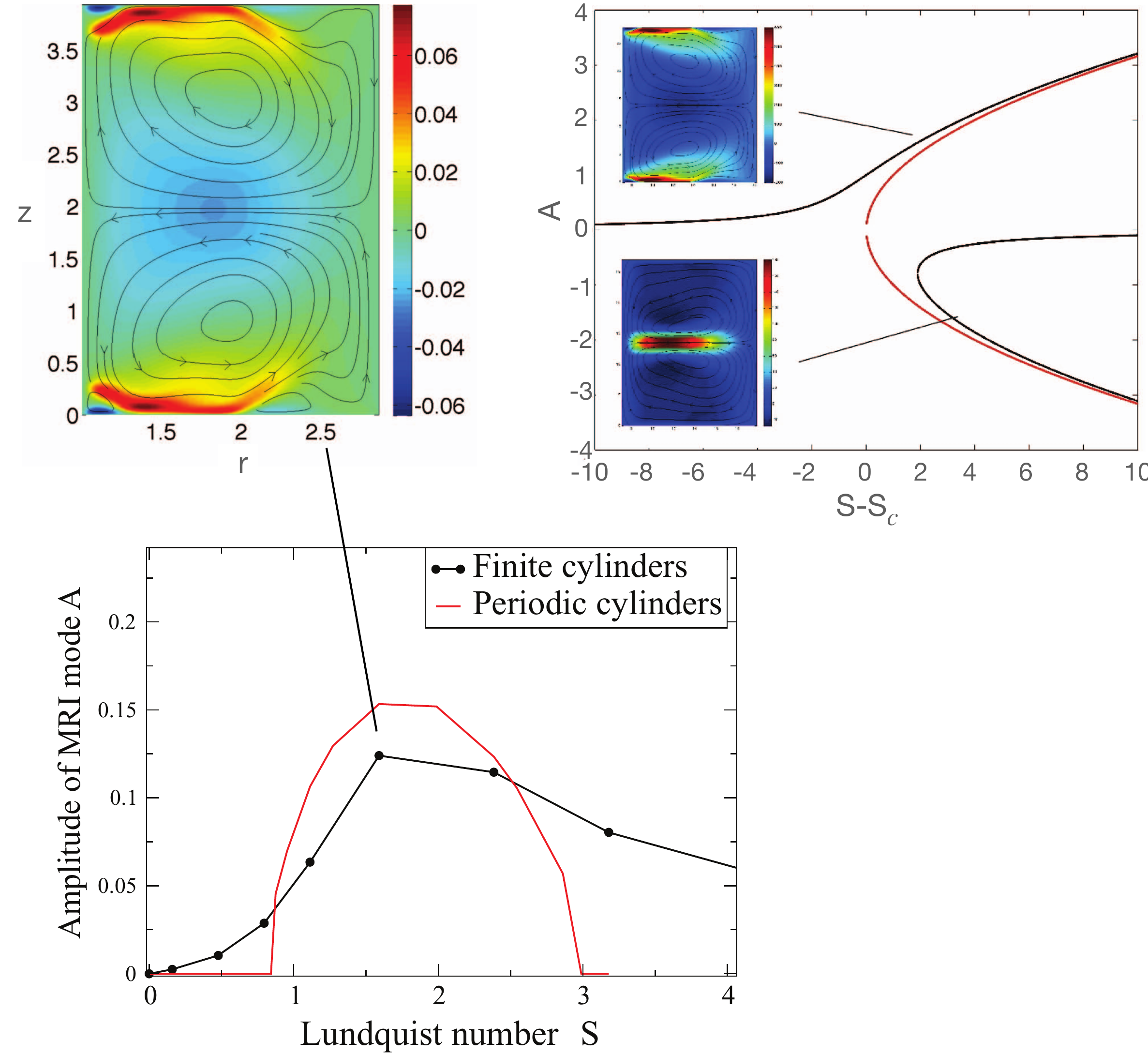}
\caption{Top-Left: structure of the MRI mode for $R_m = 15$ and $S = 2.3$ in numerical simulations~\cite{Gissinger12} showing the radial velocity $u_r$ in the ($r, z$) plane and streamlines of the poloidal flow. Bottom: bifurcation diagram of the MRI amplitude as a function of $S$ for $R_m = 15$. Red curve correspond to periodic boundary conditions in $z$, whereas the black one shows results obtained with no-slip rotating rings at the endcaps. In the latter case, the bifurcation is an imperfect supercritical pitchfork bifurcation due to the poloidal flow. Top-Right: Schematic representation of theoretical perfect (red) and imperfect (black) pitchfork bifurcations A($S-S_c$) of the MRI, where $S_C$ is the critical onset for MRI in the perfect case. Each branch corresponds to a different structure of the MRI mode.}\label{fig:bif_MRI}
\end{figure}

Numerical simulations of MRI Taylor--Couette experiments predict that the eigenmode for the MRI consists of two large poloidal recirculation cells (see Figure~\ref{fig:bif_MRI}, top-left), associated with an outward jet near the endcaps. It is therefore similar to the usual recirculation pattern generated by the presence of vertical boundaries in rotating Taylor--Couette flows. Under these conditions, how to distinguish MRI from the background flow? The answer is not easy.

To understand the difference, Figure~\ref{fig:bif_MRI}-bottom shows the bifurcation diagram of the MRI mode, for two typical simulations: when periodic boundary conditions in the axial direction are used (red curve), a perfect supercritical pitchfork bifurcation is observed. But with cylinders of finite size, there is always a non-zero poloidal recirculation generated by the endcaps, and the bifurcation is modified to a gradual transition, which does not allow to define a clear onset.\linebreak This imperfect pitchork bifurcation can be modeled~\cite{Gissinger12} by:
\begin{equation}\label{bif_eq}
{\dot A}=\mu A - A^3 +h
\end{equation}
where $h$ represents a symmetry-breaking parameter reflecting the solution's lack of axial periodicity, and which therefore prohibits the $A \rightarrow -A$ symmetry obtained in the case of periodic or infinite cylinders. Figure~\ref{fig:bif_MRI} (top-right) shows a schematic diagram of these two types of bifurcation and provides a number of predictions: Under experimental conditions, MRI emerges continuously from magnetized Ekman recirculation, and angular momentum increases progressively, well below the linear onset. On the other hand, the critical exponent $1/2$ typical for supercritical bifurcation is only measurable well above the onset predicted by linear theory. Also, different MRI structures are expected depending on the hydrodynamic background state. For instance, in the Princeton experiment, the rotation of the endcaps determines the direction of rotation of the poloidal background flow, which will select a similar structure for the MRI mode. This change in the experiment's boundary conditions corresponds to the change $h\rightarrow -h$ in equation~\eqref{bif_eq}, which selects mode $-A$. Finally, in the case of an imperfect bifurcation, the branch that is not connected to the background state is replaced by a saddle-node bifurcation, which leads to a bistability between two MRI modes with very different structures. This predicts the existence of a subcritical bifurcation towards the MRI, much easier to observe than the continuous transition generally sought~\cite{Gissinger12}.
%
%
Interestingly, although these nonlinear aspects primarily concern experimental setups, some could also apply to astrophysical disks. Any strong perturbation to the Keplerian background rotation (such as gravitational perturbation from the central object or radial suction~\cite{Gallet10} for instance) may lead to similar behavior. The existence of a subcritical transition to turbulence triggered by boundary effects is particularly appealing.

A second important nonlinear aspect involves predicting the saturated amplitude of the MRI in experimental setups. Various mechanisms have been proposed for the saturation of the MRI. In the astrophysical context, where the background shear is maintained by the gravitational field and remains unperturbed by turbulent fluctuations, a straightforward scenario suggests that MRI-driven turbulence generates sufficient turbulent dissipation to quench the MRI back to its onset. In laboratory experiments, the simplest explanation is that the instability saturates by modifying the shear responsible for its occurrence (Knobloch and Julien, 2005~\cite{Knobloch05}). More precisely, the altered shear profile produced in the bulk by the growing MRI can be balanced by an appropriate radial pressure gradient. In Taylor--Couette flows, this pressure gradient is inherently linked to the meridional recirculation induced by axial boundary conditions, thus depending on both viscous and Ohmic dissipation. Consequently, saturation in laboratory setups is closely tied to boundary effects.

Numerical modeling of magnetized Taylor--Couette flows show that in the linear phase, the growth rate is independent of the Reynolds number, underscoring the relevance of non-viscous linear theory. In contrast, the saturation phase is highly dependent on $Re$. The saturated amplitude of the magnetic perturbation $B$\ follows a scaling law $S \propto Re^{-1/4}$ (where $S=BL/(\mu_0\rho\eta)$ is the Lundquist number previously introduced) \cite{Gissinger12}, that can be traced back to the weak viscous coupling between the boundaries and the liquid metal: this scaling law can be roughly recovered by dimensional analysis as a balance between Ekman pumping ($\propto Re^{-1/2}$) and the Maxwell stress tensor in the bulk ($\propto B^2$), in agreement with the saturation mechanism proposed by Knobloch and Julien. Although such a scaling law does not directly apply to astrophysical disks, it highlights the challenge of observing MRI in the laboratory, where $Re$ typically reaches several tens of millions at the critical onset. A naive interpolation of this scaling law thus suggests an extremely low saturation amplitude for laboratory MRI. Note that a similar scaling law applies to Helical MRI, but the oscillatory nature of this instability simplifies its detection compared to the stationary standard MRI.

\section{Detection of the MRI by the Princeton experiment}

Experimental modeling of accretion disks recently took a new turn when the Princeton experiment was updated to take into account some of the remarks made above. This led the group to report convincing evidence of the experimental detection of MRI (Wang et al., 2022~\cite{Wang22}). The primary change involved implementing electrically-conducting segmented endcaps made of copper. One of the motivations was to enhance the coupling between these endcaps and the liquid metal~\cite{Wei16} thereby achieving a more powerful driving of the flow. Additionally, it appears that the modification of induced currents in the boundary layers due to the presence of these conducting endcaps alters the adverse pressure gradient. Fortunately, this change affects the scaling law $S\sim Re^{1/4}$ and increases the saturation level discussed in the previous section. This clearly increases the detectability of the standard MRI, particularly through magnetic measurements using Hall probes. But these conducting boundaries also increase the vertical extension of Shercliff layers, which significantly perturb the background flow and magnetic field. Adjustments were therefore made in the detection techniques to account for the imperfect nature of the bifurcation described in~\cite{Gissinger12}.

\begin{figure}[tbp]
\includegraphics[width=0.99\linewidth]{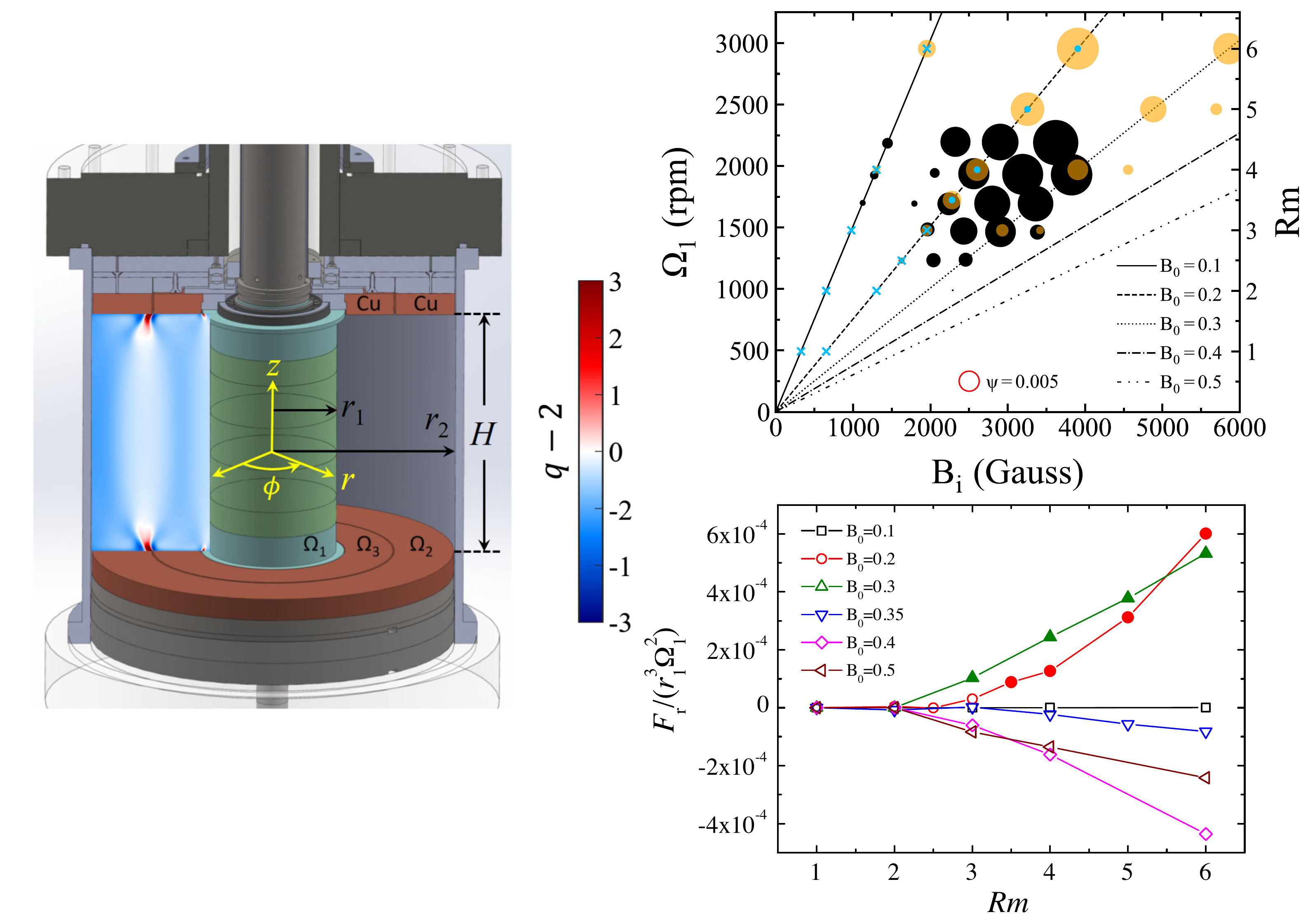}
\vspace*{-10pt}
\caption{Results obtained with the new version of the Princeton experiment (schematic view given on the left) for the Standard MRI~\cite{Wang22}. Top-right: Stability diagram in the $(\Omega_1,B)$ plane obtained in~\cite{Wang22}. The area of the circles is proportional to the magnitude of the instability and the diagram is in good agreement with the theoretical predictions of an imperfect supercritical bifurcation. Right, bottom: Evolution of the angular momentum flux with $R_m$, showing a net outward transport of angular momentum above the instability onset.}\label{fig:panel_Wang}
\end{figure}

More precisely, a radial field $B_r$ is naturally induced as soon as the vertical magnetic field $B_z$ is applied to the flow. This is represented by the parameter $h$ in equation~\eqref{bif_eq}. As expected for this type of induction process due to the residual Ekman circulation, $B_r$ was reported to evolve linearly with $Rm$. However, beyond a critical onset $Rm_c$, a nonlinear increase in $B_r$ is observed and interpreted as a signature of the MRI, as predicted by equation~\eqref{bif_eq}. The instability thus exhibits all the characteristics of an imperfect pitchfork bifurcation typical of MRI (see Figure~\ref{fig:panel_Wang}): it is unstable only beyond a specific magnetic Reynolds number $R_m$ (well above the onset of inductionless instabilities), it transitions gradually from residual Ekman recirculation, and restabilizes if the magnetic field becomes too strong, all while presenting an axisymmetric eigenmode consistent with numerical predictions. Moreover, the experiment confirmed a weak but clear outward transport of angular momentum, attributed to both velocity fluctuations and the magnetic torque induced by the MRI instability, corroborating theoretical expectations of MRI in accretion disks. It has also been checked that this instability is not related to the Shercliff layer instabilities reported previously.

An unexpected observation is the detection of non-axisymmetric MRI modes alongside the $m=0$ mode, suggesting a complex interplay of MRI modes with different symmetries~\cite{Wang22b}. This complexity warrants further investigation, as it may provide insights into how MRI could lead to fully developed turbulence through nonlinear interactions between multiple modes.

\section{Beyond Taylor--Couette MRI: angular momentum transport}

\subsection{Limitations inherent to Taylor--Couette devices}\label{limita}

The recent success of the Princeton MRI experiment, along with the alternative scenarios proposed by the PROMISE experiment, has significantly advanced our understanding of accretion disks. These developments now provide a viable scenario for Keplerian flows to become unstable and conclusively demonstrate that the magnetorotational instability can be generated in a real system. This represents a substantial progress in unraveling the destabilization of astrophysical disks, a puzzle that has persisted for at least five decades. To draw definitive conclusions, however, a complementary approach beyond the Taylor--Couette geometry is necessary. Taylor--couette devices present at least three significant limitations for accurately modeling accretion disks:
\begin{itemize}
\item[-] Firstly, the velocity profile in these experiments, while quasi-Keplerian, differs from a true Keplerian rotation rate. In other words, the radial stratification of angular momentum is different, leading to turbulent transport properties that can be different from those of accretion disks.

\item[-] Secondly, even when quasi-Keplerian flows become unstable to MRI, the resulting turbulence remains very weak. The reason is that it is exceedingly difficult to increase the magnetic Reynolds number beyond the instability threshold, resulting in only minimal angular momentum transport. This necessity of observing Keplerian turbulence in the laboratory is important, especially since recent observations have suggested that the level of turbulence in Keplerian disks may be much smaller than expected before~\cite{Flaherty15}.

\item[-] Thirdly, and perhaps most critically, Taylor--Couette devices rely on rotating boundaries, which corresponds to a surface injection of angular momentum: it is produced on the rotating vessel walls, then diffuse through the viscous boundary layers located on the cylinders, to finally produce transport in the turbulent bulk. This is a scenario far removed from that of astrophysical disks, which are essentially immune to boundary effects, especially for angular momentum transport.
\end{itemize}

\subsection{Plasma experiments}\label{sec_plasma}

In this perspective, alternative approaches are necessary. One possibility is the use of plasma. An active research effort is currently underway to develop Taylor--Couette equivalents in plasmas, where the flow is instead generated by a series of electrodes injecting currents through the plasma~\cite{Collins12, Flanagan20}.

Two experiments of this type are currently being developed at the Wisconsin Plasma Physics Laboratory: the 1-meter-wide cylindrical Plasma Couette Experiment (PCX) and the 3-meter-wide spherical Big Red Ball (BRB). In the PCX experiment, an argon plasma is generated at approximately 10,000 K by a hot lanthanum hexaboride (LaB6) cathode located at the top of the cylindrical axis and collected by cold molybdenum anodes positioned near the outer equatorial edge. A radial current flows through an externally applied vertical magnetic field, creating a torque that drives the plasma in the toroidal direction, producing a Couette velocity profile. While an array of permanent magnets on the boundary helps confine the plasma, the bulk magnetic field remains weak (typically $\le 10$ G). The BRB experiment operates on a similar principle but with reversed electrode positions—the molybdenum anodes are placed at the top axis. The plasma in these experiments operates in a flow regime quite distinct from the classical MHD framework relevant to liquid metals. Here, the ratio $\beta$ between kinetic and magnetic pressures is large, and plasma dynamics occur on scales smaller than the ion inertial length ($\approx 1$ m here) but larger than the electron inertial length. At these scales, ions and electrons are decoupled, requiring a two-fluid description. Electrons remain tied to the magnetic field, while the ion fluid decouples from it. This decoupling is accounted for by adding the Hall term to Ohm's law which now becomes:
\begin{equation}
\bfJ=\sigma\left(\bfE + {\bfv\times B}-\frac{1}{n_ee}\left({\bfJ\times B} \right)\right)
\end{equation}
where $n_e$ is the electron number density, $e$ is the elementary charge, and the last term is the Hall term. The plasma is therefore described by what is known as the collisionless Hall regime.
In~\cite{Flanagan20}, it has been shown that a significant amplification of the magnetic field (by a factor of 20) occurs when the electrons flow radially inwards, while an almost total expulsion of the field is observed when the flow is reversed - behaviour that is well explained by the Hall effect. Plasma Couette flow experiments are crucial as they expand MRI research to include Hall effects and kinetic physics. These effects directly address key issues relevant to hot and dense disks, and partially magnetized protostellar accretion disks~\cite{Kunz13}. Beyond MRI, the Hall effect has significant applications, including turbulent reconnection in magnetospheres~\cite{Phan18} and the evolution of magnetic fields in neutron star crusts~\cite{Cumming04}.

\begin{figure}[tbp]
\includegraphics[width=0.99\linewidth]{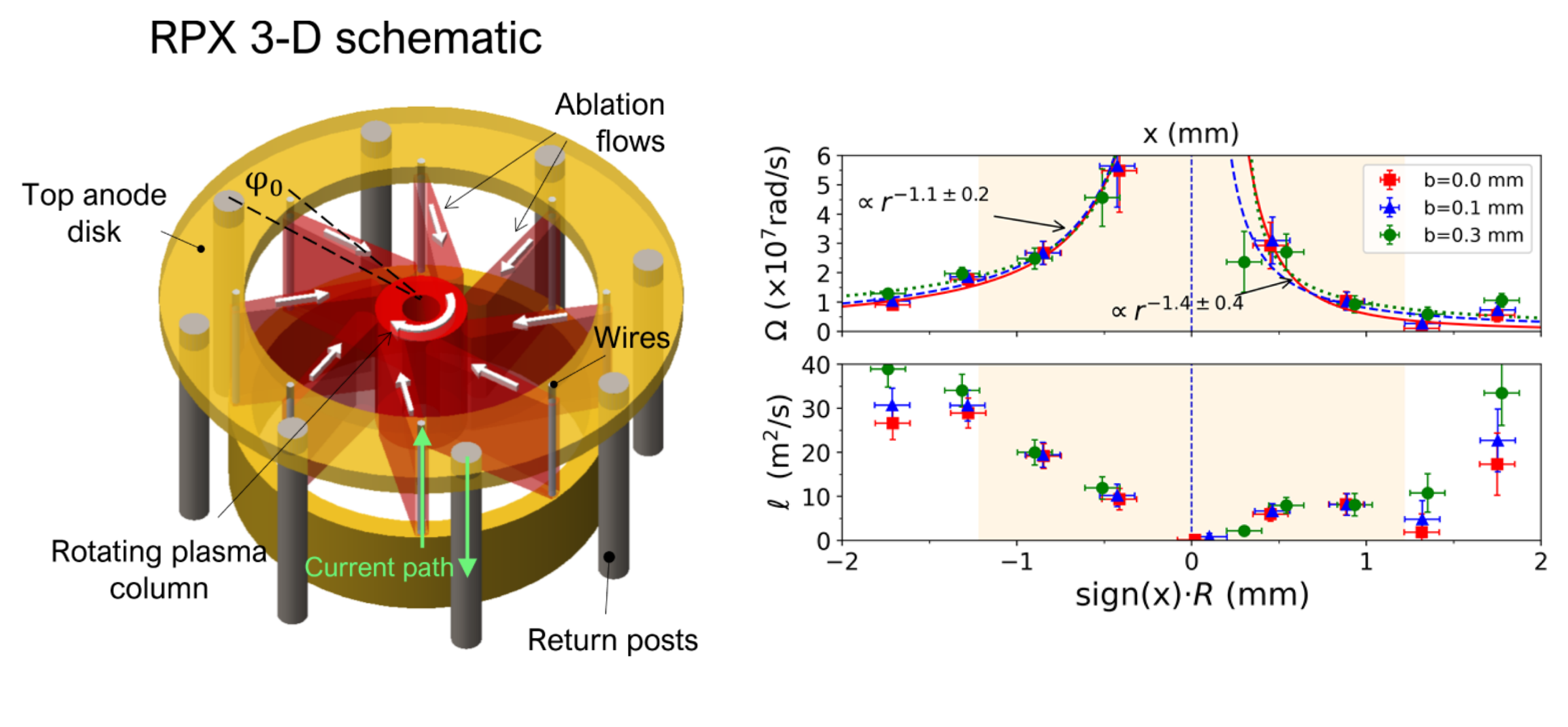}
\caption{Left: Schematic diagram of the experiment using an inertially driven rotating plasma reported in~\cite{Valenzuela23}. The setup uses the oblique collision of multiple plasma jets, resulting from the ablation flows from a cylindrical aluminum wire array (eight $40 \mu$m wires, $16$ mm diameter, $10$ mm high), to inject both mass and angular momentum into the rotating plasma. Right: radial profile of both the angular roation $\Omega$ (top) and the angular momentum distribution (bottom), showing quasi-Keplerian flows.}
\label{fig:panel_plasma}
\end{figure}

Another recent laboratory study~\cite{Valenzuela23} was proposed in this direction by reporting on a pulsed-power driven differentially rotating plasma.
In this setup, a strong current flows axially through a cylindrical array of aluminum wires (called a Z-pinch configuration in reference to the axial direction Z), causing the wires to heat up resistively and generate a low-density plasma by ablation. The central cathode directs the return current along the axis of the array, driving ablation flows radially. The ram pressure resulting from these ablation flows induces a rotating plasma column near the center, effectively injecting angular momentum without relying on boundary-driven methods (see Figure~\ref{fig:panel_plasma}). The rotational speed profile is quasi-Keplerian, and the flow rotates subsonically, with an azimuthal velocity of $23$ km.s$^{-1}$, a rotational speed in no way comparable to flows of a few $m/s$ obtained in liquid metals. This translates into Reynolds numbers of the order of $Re\sim10^4$ and $R_m\sim 30$. It should be noted, however, that it would not be so easy to reach the onset of MRI, as the fastest MRI growth mode would develop on timescales of the order of $100$ ns and for a critical magnetic field of up to $5$ Tesla depending on plasma density. These experiments nevertheless represent a key initial step in establishing a laboratory plasma framework for basic investigations of accretion disks physics.

\subsection{The Kepler experiment}

The Kepler experiment at ENS Paris is another approach, which avoids the measurements difficulties associated with laboratory plasmas. It was also designed specifically to address the three limitations discussed in Section~\ref{limita}, and to focus on the angular momentum transport observed in Keplerian turbulence (Vernet et al., 2021~\cite{Vernet21a},Vernet et al., 2022~\cite{Vernet22}). The primary aim of this experiment is not to demonstrate MRI or any other mechanism for destabilizing accretion disks, but rather to study the transport properties of Keplerian turbulence in thin disks, independently of the exact mechanisms that generate this turbulence.

In this experiment, the magnetic Reynolds number is less than one, indicating that induced magnetic fluctuations are negligible compared to the applied magnetic field. In this Quasi-Static limit, neither the Magneto-Rotational Instability (MRI) nor dynamo instabilities are expected, a situation close to the dead zones found in real accretion disks. On the other hand, a very large turbulent angular momentum transport is obtained in a Keplerian flow. It is therefore an essential complementary approach to the Taylor--Couette MRI experiments.

\begin{figure}[tbp]
\includegraphics[width=0.99\linewidth]{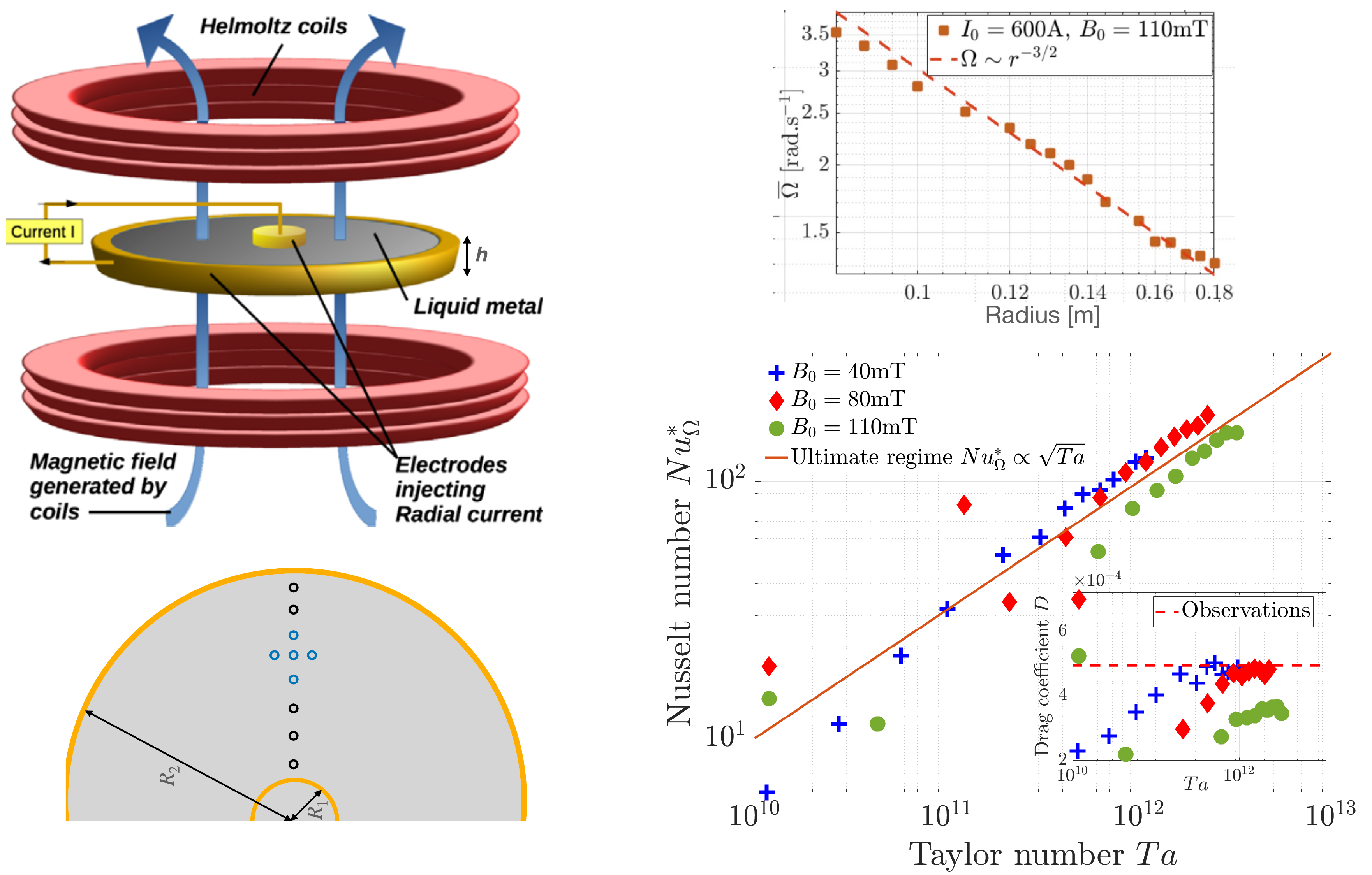}
\vspace*{-5pt}
\caption{Left: Setup of the Kepler experiment~\cite{Vernet21a,Vernet22}. A thin disk (height $h=15$ mm, diameter $D=380$mm) filled with liquid galinstan is placed between two Helmholtz coils applying a vertical magnetic field. To drive the flow, radial boundaries are not rotating, but acts as electrodes injecting a radial current through the liquid metal. A series of potential probes give access to radial and azimuthal velocity in the bulk. Right-top: Keplerian velocity profile $\Omega\sim r^{-3/2}$ measured in the bulk. Right-bottom: Scaling law $Nu_\Omega$ vs $Ta$ reported in the Kepler experiment~\cite{Vernet22}, showing good agreement with Kraichnan's prediction of an ultimate angular momentum transport by Keplerian turbulence. Inset: dimensionless drag coefficient compared to observations (see text).}
\label{fig:kepler_setup}
\vspace*{-10pt}
\end{figure}

The setup includes two concentric cylinders that mimic the thin geometry of a disk, measuring 380 mm in diameter and only 15 mm in height (see Figure~\ref{fig:kepler_setup}). Insulating Plexiglass endcaps seal the system, which is filled with Gallinstan and placed between two large Helmholtz coils that generate a vertical magnetic field. Unlike previous Taylor--Couette experiments, the radial boundaries here do not rotate. Instead, the inner and outer cylinders are brass Nickel-plated electrodes through which a strong radial electrical current (approximately $I_0\sim 3000$ Amps) is applied. This current, combined with the vertical magnetic field $B_0$ generated by the coils, generates a volumetric Lorentz force that drives the liquid metal in the azimuthal direction. Velocity fields in both radial and azimuthal directions are measured through ultrasound Doppler velocimetry and a series of potential probes.

Figure~\ref{fig:kepler_setup} shows that a critical difference from previous experiments is the angular rotation generated in the bulk flow, much closer to that obtained in astrophysics. Except near the boundaries where the velocity must go to zero, the Lorentz force, which replaces gravitation in the force balance with fluid's inertia, imposes an exact Keplerian rotation profile $\Omega \sim r^{-3/2}$, for the first time observed in the laboratory.

The situation is not, however, comparable to the first phase of a Keplerian disk, before the MRI or another instability destabilizes the flow. On the contrary, this velocity profile is turbulent, and only appears at very high forcing, after a series of centrifugal and boundary layers instabilities~\cite{Vernet21a} have occured. Consequently, the Kepler experiment models the final state of an accretion disk, which exhibits the same Keplerian turbulence once destabilized by MRI or any other instability. This approach has a clear advantage: by relying on a transition to turbulence easier to obtain than MRI, it allows us to reach a more turbulent state, and thus to study for the first time the transport properties of Keplerian turbulence in a thin disk.

The generation of Keplerian turbulence by the Lorentz force also provides a volume injection of angular momentum, a more satisfactory mechanism compared to the boundary injection achieved by rotating cylinders in Taylor--Couette geometries. Using local measurements of radial and azimuthal velocity fields, \cite{Vernet22} reported the flux of angular momentum $J_\Omega=r^3(\overline{u_r\Omega}-\nu\partial_r\overline{\Omega})$, as a function of the fluid's rotation rate over several orders of magnitude. This transport can be quantified by two dimensionless numbers: the Nusselt number $ Nu = J_\Omega / (2 \nu \Omega r^2)$, which measures the efficiency of angular momentum transport relative to the laminar case, and the Taylor number $ Ta = 4r^2u_\varphi^2/\nu^2$, which quantifies the angular rotation of the fluid.

Two primary contributions to local angular momentum transport were identified: one from the poloidal recirculation induced by the boundaries, similar to the one observed in Taylor--Couette devices, and a contribution from turbulent fluctuations in the bulk flow. Figure~\ref{fig:kepler_setup} demonstrates that for sufficiently large Taylor numbers, the experimental data align well with the asymptotic scaling law:

\begin{equation}\label{eq:ultimate_regime}
Nu_\Omega^* \propto \sqrt{Ta}
\end{equation}
where the star indicates that the contribution from poloidal circulation has been subtracted. The observation of this law is a significant achievement as it represents a scaling known for describing efficient angular momentum transport independent of the fluid's molecular viscosity. This scaling was predicted by Kraichnan 60 years ago in the context of heat transport by astrophysical turbulence, and represents the most efficient regime for angular momentum transport by a turbulent flow. This so-called ``ultimate'' regime of transport had not been reported in previous Taylor--Couette experiments: in those setups, as mentioned above, the angular momentum injected at the boundaries systematically diffuses through the thin viscous layer on the cylinders. This diffusion in the boundary layer dominates much of the transport, and it is impossible to avoid the effect of molecular viscosity, thus preventing the emergence of a diffusivity-free angular momentum regime. In contrast, the Kepler experiment circumvents this constraint by injecting angular momentum volumetrically via the Lorentz force and reaching a state in which the boundary layers are fully turbulent, thus suppressing the role of the molecular viscosity on the transport.
 This approach is remarkably similar to the method used by Bouillaut et al., 2019~\cite{Bouillaut19} for thermal convection. In their experiment, turbulent thermal convection was generated by applying radiative forcing to the bulk rather than heating the boundaries. This technique produced highly efficient heat transport in the flow, bypassing the boundary layers and reaching the ultimate regime of turbulent convection. Although focused on different topics, these two experiments echo each other as distinct manifestations of the non-molecular ultimate regime predicted by Kraichnan.

This asymptotic scaling law is particularly relevant for astrophysical systems characterized by low viscosity and negligible boundary layers, and is sometimes presented as a universal prediction for astrophysical transport. Indeed, it can be recovered for the Kepler experiment using just a few general arguments: firstly, the angular momentum flux $J_\Omega$ must be equal to the angular momentum rate injected locally by the Lorentz force, $J_\Omega\sim I_0B_0r^2/(2\pi\rho h)$ (which can be seen as a consequence of the Keplerian rotation produced by the magnetic forcing). Secondly, a simple energy balance dictates that the rate of energy dissipation $\varepsilon$ must be balanced by the work of the Lorentz force driving fluid motion, i.e. $\varepsilon\sim \int_V(\bm{j \times B}).\bm{u} \propto I_0B_0u_\varphi$. Combining these two results, we obtain $\varepsilon= \gamma J_\Omega u_\varphi$, with $\gamma$ a geometric parameter of order one. Finally, Kraichnan's argument of an ultimate regime without molecular diffusivity leads to the expression $\varepsilon\sim u_\varphi^3/\ell$ where $\ell$ is the appropriate mixing length and we have assumed that radial boundary layers are fully turbulent. In the case where the mixing length $\ell$ is chosen to be the disc thickness $h$, the scaling law~\eqref{eq:ultimate_regime} is finally recovered, with the correct prefactor. This simple calculation also shows that the transport properties are relatively independent of the details of the instability driving the turbulence, an important point given recent doubts about the applicability of MRI to cold accretion disks. Here the only three important assumptions are a molecular-free transport, a Keplerien rotation and some specification of the mixing length.

Despite substantial differences between this experiment and real accretion disks, it thus enables predictive analysis. In~\cite{Dubrulle05,Hersant05} it was shown that a dimensionless drag $D$ can be computed from observations -- it is proportional to the accretion rate of protoplanetary disks and correlates directly with the $\alpha_{SS}$ parameter. Interestingly, this dissipation coefficient $D$ can also be determined in laboratory experiments from the $Nu_\Omega^*(Ta)$ scaling law, and is given by $D= 4Nu_\Omega^* / \sqrt{Ta}$. The inset of Figure~\ref{fig:kepler_setup} shows that this dimensionless turbulent drag simply tends to a constant value of $D\sim 5\times 10^{-4}$ in the Kepler experiment, as expected in the ultimate regime where $Nu_\Omega^*/\sqrt{Ta} = \text{constant}$. This is an advantage compared to Taylor--Couette devices, where $D$ varies with the Reynolds number and cannot be directly compared to real accretion disks operating at extremely high $Re$ values. This constant value $D$ closely matches the one measured in disks around T Tauri stars: the median value of the observational data reported by~\cite{Hersant05} is $4,9\times 10^{-4}$, as indicated by the red dashed line in the inset of Figure~\ref{fig:kepler_setup}).

This remarkable agreement warrants careful interpretation. While the median value of $D$ from observations aligns with the Kepler experiment, calculating the mean value yields $D_{obs} \sim 2 \times 10^{-3}$, indicating that many accretion disks exhibit significantly higher values than those observed in the Kepler experiment. However, it is important to note that the Kepler experiment operates in a parameter regime quite distinct from real accretion disks, particularly with a magnetic Reynolds number below unity (compared to $Rm \sim 10$ in the Princeton experiment). Consequently, the transport observed in the Kepler experiment is purely due to hydrodynamical stress, without any contribution from Maxwell stress, which would be expected in the presence of MRI. The simple calculation above suggests that the experimental value of $D$ is consistent with a scenario of pure hydrodynamical transport, characterized by non-molecular angular momentum transport, Keplerian turbulence, and a mixing length based on disk thickness. Deviations from these conditions will lead to different values of $D$. In other words, the Kepler experiment and its value $D \sim 5 \times 10^{-4}$ likely represents a minimal bound for ultimate turbulent transport in Keplerian disks, particularly in the absence of Maxwell stress, as expected in cold disks or dead zones. Any observational value exceeding this number could however be interpreted as an indication of MRI. This experimental study could thus be extremely useful for a quantitative analysis of observational data and their interpretation in terms of turbulence models.

\section{Conclusion and perspectives}

This article presents a comprehensive review of laboratory studies aimed at modelling the dynamics of accretion disks, focusing on magnetohydrodynamic experiments carried out over the last two decades. Taylor--Couette devices, such as the one developed at the Princeton University, have focused on the destabilization of Keplerian disks. It has been shown that even at high Reynolds numbers, Keplerian or quasi-Keplerian flows appear to remain stable without MHD effects. However, when an external magnetic field is applied and for sufficiently large inductive effects, a stable centrifugal flow can become unstable to MRI. It then follows a scenario relatively close to the one relevant to astrophysical disks, exhibiting outward angular momentum transport and complex turbulent fluctuations. This remarkable result was only possible after a long journey through the role of boundary conditions and nonlinearities in rotating flows.

Laboratory modeling of astrophysical disks also led to the discovery of other MHD instabilities such as the Shercliff layer instability, or the one reported by the group in Dresden: the Promise experiment have reported the discovery of a new type of MRI in which the presence of an azimuthal field strongly modifies the flow dynamics, producing a magnetorotational instability at very small $R_m$. This approach opens up interesting new avenues for the weakly ionized regions of disks for which SMRI could not operate.

Finally, these studies of the first stage of an accretion disk, focusing on the origin of turbulence, have recently been complemented by an experimental model of the final stage of the accretion disk, which reproduces the Keplerian turbulence of a thin disk in the laboratory. Providing a quantitative study of angular momentum transport by astrophysical turbulence, the Kepler experiment offers new constraints for turbulent transport in accretion disks, universal enough to be applied to a large range of proto-planetary disks, MRI-unstable or not.

Looking ahead, and building on the recent success of these experiments, we have a number of options open to us to address some outstanding questions on astrophysical disks.

An immediate improvement, currently ongoing in the Princeton experiment, is to make additional measurements in the MRI-unstable regime. Improved flow characterization, for example using multidimensional ultrasonic Doppler velocimetry and Hall probe arrays, will enable detailed measurements of the various components of velocity and magnetic fields. In particular, analysis of their correlations would greatly help to understand the transport of angular momentum by the instability.

In light of the discussions in previous sections, another question that emerges relates to the nonlinear dynamics of MRI. Currently, most results suggest that nonlinear effects appear to be system-dependent, influenced by factors such as geometry, materials, and boundary conditions, and a generic understanding of these aspects is still lacking~\cite{Mishra23}. In particular, a clear understanding of the mechanisms controlling the saturation of MRI and the corresponding route to fully developed turbulence in laboratory experiments would be of great interest, both for fundamental physics and accretion disk studies. To achieve this, future experiments need to explore the instability well beyond its onset, an exciting challenge due to the considerable power requirements.

In Dresden, several new MHD experiments are currently developed in order to extend the results obtained from the PROMISE experiment~\cite{Stefani19}. It includes a new version of the Taylor--Couette experiment using liquid sodium, which is a much better electrical conductor than gallium. This enhancement opens up new opportunities. One possibility is to test the MRI-driven dynamo scenario, in which flow perturbations generated by the MRI can produce their own magnetic field through dynamo action. This scenario is extremely popular in astrophysics because it provides a self-sustaining mechanism to explain both turbulence and the magnetic field that sustains the MRI in accretion disks.

We also mentioned several attempts to study this problem using plasma experiments, This is a very promising approach since many limiting factors associated with liquid metals disappear. The magnetic Prandtl number $P_m$ influences various aspects of MRI such as the linear onset, the saturation and the level of fluctuations. In plasma, the viscosity can be varied independently of the conductivity, allowing $P_m$ to range from values typical of liquid metals ($P_m \ll 1$) to those akin to interstellar plasmas ($Pm\gg 1$), an interesting tool for understanding the nature of plasma turbulence~\cite{Collins12}. In addition, plasma experiments operating at $Pm\sim 1$ offer the possibility of reaching high magnetic Reynolds numbers while maintaining a kinetic Reynolds number that is not excessively large. This is a significant advantage for observing many MHD instabilities, which are generally known to exhibit a substantial increase of the onset with the level of turbulence.

Finally, by increasing both the size of the disk and the conductivity of the fluid in the Kepler experiment, the Keplerian flow may also become MRI-unstable.
Interestingly, this possibility of studying MRI with an electromagnetically-driven flow of liquid sodium was proposed by Velikhov himself~\cite{Velikhov06}. Analytical and numerical calculations support this~\cite{Khal06,Khalzov10}, suggesting that the critical onset of the MRI could be reached with a larger version of the Kepler experiment, provided the aspect ratio $h/R$ is increased. An ongoing modification of the experiment involves filling this larger setup with two non-miscible fluids, specifically a free-surface layer of liquid gallium atop a layer of mercury. This layered configuration offers two major advantages: first, it significantly reduces top and bottom friction, allowing the gallium layer to rotate at much faster rates while remaining centrifugally stable, a situation highly conducive to the detection of MRI. Second, by carefully selecting the thickness of these layers, the experiment can be conducted in the shallow water limit. Invoking the analogy between shallow water dynamics and compressible gas dynamics~\cite{Gunzkofer21,Foglizzo12}, this type of setup could provide the first laboratory study of an accretion disk incorporating compressible effects.

These different examples illustrate how a set of complementary approaches, with diverse experimental setups, can collectively improve our understanding of accretion disks. However, such advances will only be possible in close collaboration with observational data. At a time when observational techniques are undergoing revolutionary changes, we could not hope for a better opportunity to deepen our understanding of astrophysical disks.

\section*{Declaration of interests}
The authors do not work for, advise, own shares in, or receive funds from any organization that could benefit from this article, and have declared no affiliations other than their research organizations.





%
%
%
%
%
%
%
%

\bibliographystyle{crunsrt}


\bibliography{review_GAFD_disks_bib}

\end{document}